\documentclass[lettersize,journal]{IEEEtran}
\usepackage{amsmath,amsfonts}
\include{pythonlisting}
\usepackage{algorithmic}
\usepackage{algorithm}
\usepackage[algo2e,ruled,vlined]{algorithm2e}
\usepackage{array}
\usepackage{textcomp}
\usepackage{stfloats}
\usepackage{url}
\usepackage{verbatim}
\usepackage{graphicx}
\usepackage{cite}
\usepackage{url} 
\usepackage{soul}
\usepackage{multirow}
\usepackage{nicematrix}
\usepackage{subfigure} 
\usepackage{xcolor}
\usepackage{colortbl}

\begin{document}

\title{Smart Mobility Digital Twin Based Automated Vehicle Navigation System: A Proof of Concept}

\author{Kui~Wang,~\IEEEmembership{Student Member,~IEEE,}
      Zongdian~Li,~\IEEEmembership{Member,~IEEE,}
      Kazuma~Nonomura,~\IEEEmembership{Student Member,~IEEE,}
      Tao~Yu,~\IEEEmembership{Member,~IEEE,}
      Kei~Sakaguchi,~\IEEEmembership{Senior Member,~IEEE,}
      Omar~Hashash,~\IEEEmembership{Student Member,~IEEE,}
      and~Walid~Saad,~\IEEEmembership{Fellow,~IEEE}
      
\thanks{This work was supported by the Japan National Institute of Information and Communications Technology (NICT) under JUNO Grant 22404, the U.S. National Science Foundation under JUNO3 Grant CNS-2210254, the Japan Science and Technology Agency (JST) under SPRING Grant JPMJSP2106, and by the Tokyo Tech Academy for Super Smart Society (SSS). (\textit{Corresponding author: Zongdian Li, Kui Wang.})}

\thanks{K. Wang, Z. Li, K. Nonomura, T. Yu, and K. Sakaguchi are with the Department of Electrical and Electronic Engineering, School of Engineering, Tokyo Institute of Technology (Email: \{kuiw, lizd, nonomura, yutao, sakaguchi\}@mobile.ee.titech.ac.jp).}

\thanks{O. Hashash and W. Saad are with the Bradley Department of Electrical and Computer Engineering, Virginia Tech, Arlington, VA 22203 USA (Email: \{omarnh, walids\}@vt.edu).}}



\maketitle
\thispagestyle{plain} 

\begin{abstract}
Digital twins (DTs) have driven major advancements across various industrial domains over the past two decades. With the rapid advancements in autonomous driving and vehicle-to-everything (V2X) technologies, integrating DTs into vehicular platforms is anticipated to further revolutionize smart mobility systems. In this paper, a new smart mobility DT (SMDT) platform is proposed for the control of connected and automated vehicles (CAVs) over next-generation wireless networks. In particular, the proposed platform enables cloud services to leverage the abilities of DTs to promote the autonomous driving experience. To enhance traffic efficiency and road safety measures, a novel navigation system that exploits available DT information is designed. The SMDT platform and navigation system are implemented with state-of-the-art products, e.g., CAVs and roadside units (RSUs), and emerging technologies, e.g., cloud and cellular V2X (C-V2X). In addition, proof-of-concept (PoC) experiments are conducted to validate system performance. The performance of SMDT is evaluated from two standpoints: \textit{(i)} the rewards of the proposed navigation system on traffic efficiency and safety and, \textit{(ii)} the latency and reliability of the SMDT platform. Our experimental results using SUMO-based large-scale traffic simulations show that the proposed SMDT can reduce the average travel time and the blocking probability due to unexpected traffic incidents. Furthermore, the results record a peak overall latency for DT modeling and route planning services to be 155.15~ms and 810.59~ms, respectively, which validates that our proposed design aligns with the 3GPP requirements for emerging V2X use cases and fulfills the targets of the proposed design. Our demonstration video can be found at https://youtu.be/3waQwlaHQkk.

\end{abstract}

\begin{IEEEkeywords}
smart mobility digital twin, navigation system, vehicle-to-everything, cloud and edge computing, implementation
\end{IEEEkeywords}

\section{Introduction}
\subsection{Background}
\IEEEPARstart{D}{igital} twins (DTs) enable a synergistic integration between the physical and cyber worlds, thereby becoming a catalyst for the ongoing digital transformation of our society \cite{kritzinger2018digital}. In essence, the continuous evolution of DTs over the past two decades signifies their underlying potential to lead this digital transformation \cite{vanderhorn2021digital}. In particular, DTs possess remarkable abilities to establish dynamic digital models of the environment and bidirectional communications between the physical and cyber worlds. Thereby, many functionalities such as real-time monitoring, design, and optimization can be developed with DTs acting as their founding basis. The remarkable advancements have driven revolutionary applications into various sectors like manufacturing, agriculture, and transportation \cite{tao2018digital, thomas2023causal}. 

At the forefront of these viable sectors, integrating DTs into the transportation and traffic fields constitutes a promising solution for critical challenges like traffic congestion and road accidents \cite{rudskoy2021digital}. A recent report indicates that the global DT market reached USD 11.12 billion in 2022 and was predicted to grow at a compound annual growth rate (CAGR) of 37.5\% until 2030 \cite{dtmarket}. Notably, the automotive and transport segment has dominated the DT market in 2022, accounting for over 20\% of its total revenue. This remarkable market performance highlights the crucial role of DTs as a cornerstone in the smart mobility infrastructure \cite{sevenworld, yu2023internet}. 

Indeed, the convergence of smart mobility systems with DTs has ushered in new promising levels of autonomous driving breaking into connected and automated vehicles (CAVs), thereby introducing the concept of parallel intelligence \cite{han2023parallel, yang2023resource}. On the one hand, having these vehicles ``connected" underpins the feasibility of DT deployment. In particular, vehicles outfitted with vehicle-to-everything (V2X) can share information with various external entities by using vehicle-to-vehicle (V2V), vehicle-to-infrastructure (V2I), and vehicle-to-cloud (V2C) communications \cite{sakaguchi2021towards}. In particular, V2C facilitates vehicle data transmission to the cloud, naturally enabling the establishment of DTs on the cloud plane \cite{wang2022mobility}. On the other hand, the performances of onboard autonomous driving systems are hindered by multiple constraints, such as onboard sensing view and computing resources \cite{hobert2015enhancements}. Henceforth, incorporating DT technologies or parallel intelligence is anticipated to overcome such challenges that have hindered the evolution of CAVs.

Despite these promising strides, a significant gap persists in validating these advantages upon integrating DTs in an end-to-end (E2E) real-world smart mobility. Indeed, the theoretical advantages of DTs, as demonstrated in \cite{omar2023towards, hu2023how, lin2023mobility50}, still require practical validation to ascertain their feasibility and impact. the overarching goal of this paper is to propose a novel smart mobility DT (SMDT) platform that can be used to establish a real-world and real-time traffic DT, to enhance CAVs’ ability to ``see more and see further" in autonomous driving scenarios.

\subsection{Related Works}

\begin{table*}[ht]
\centering
\caption{Feature Comparison of Related Works}
\label{tab: compare}
\begin{tabular}{c|ccc|c|cc|c|c}
\hline
\multirow{2}{*}{Ref.}                       & \multicolumn{3}{c|}{Key components} & \multirow{2}{*}{\begin{tabular}[c]{@{}c@{}}SAE\\level \end{tabular}} & \multicolumn{2}{c|}{DT functions} & \multirow{2}{*}{\begin{tabular}[c]{@{}c@{}}Validation methods\end{tabular}} & \multirow{2}{*}{Objectives (use cases)}              \\
                                            & Cloud      & vehicles   & RSUs      &                                                                               & Modeling         & Service        &                                                                                &                                                      \\ \hline
{[}12{]}     & $\surd$    & $\surd$    & $\surd$   &   L2                                                                          & $\surd$          & $\surd$        & Field trial\&Simulation                                                                     & Personalized adaptive cruise control (P-ACC). \\
{[}25{]}   & $\surd$    & $\surd$    & $\surd$   &  L2                                                                           & $\surd$          &                & Field trial                                                                    & Cooperative perception to create traffic DT.  \\
{[}26{]}            & $\surd$    &            & $\surd$   &                                                            & $\surd$          &                & Field trial                                                                   & Cooperative vehicle-infrastructure system.           \\
{[}29{]}  &            & $\surd$    &           & L4                                                                          & $\surd$          &                & Field trial                                                                    & Validation framework for autonomous driving.         \\
{[}33{]}    & $\surd$    & $\surd$    &           &                                                                             &                  & $\surd$        & Simulation                                                                     & Path planning considering unexpected events.  \\
{[}34{]}       & $\surd$    & $\surd$    & $\surd$   &                                                                              & $\surd$          & $\surd$        & Simulation                                                                     & Traffic light control system.                        \\
{[}37{]}       & $\surd$    & $\surd$    &           &  L2                                                                            & $\surd$          & $\surd$        & Simulation                                                                     & Cooperative driving at intersections. \\
{[}38{]}        & $\surd$    & $\surd$    &           &                                                                      & $\surd$          &                & Sandbox                                                                        & Validation of multi-vehicle cooperation.             \\ \hline
{This work}        & $\surd$    & $\surd$    & $\surd$   & L4                                                                       & $\surd$          & $\surd$        & Field trial\&Simulation                                                                       & CAV Navigation system             \\ \hline
\end{tabular}
\end{table*}

As discussed above, given the benefits of DT technology, researchers have made some efforts to introduce the concept of DTs to vehicular technologies in recent years \cite{zheng2023opencda, ma2024driver, wang2023cosimulation, liang2023shared}. To build a real-time DT model of traffic environments and provide feedback services for CAVs, it is important to handle three technical challenges: sensing, communication, and computation. Sensing plays a pivotal role in environmental information capture, which is integral for the accurate representation and predictive analysis of complex vehicular and traffic systems. The work in \cite{yue2021evolution} summarizes different ways for traffic information acquisition and traffic congestion monitoring, such as using stationary roadside sensors, probe-vehicle-based techniques, vehicular networks, and social networks. Among these various methods, stationary roadside sensors, typically referred to as roadside units (RSUs), can accurately and continuously monitor the traffic condition and traffic flow \cite{cui2019automatic, wang20243dlidar}. Although the construction of RSUs is currently expensive due to the inclusion of some specialized and high-performance hardware like LiDARs, such cost can significantly decrease while the hardware becomes mass-market and widely adopted in autonomous driving and infrastructures of smart cities \cite{degrande2021c}. Hence, the use of sensor-equipped RSUs to establish a traffic DT has become popular in some related research. For example, in \cite{tihanyi2021towards}, multiple sensors deployed in the smart infrastructure are used to detect pedestrians and realize object-level central perception on the cloud. The authors in \cite{guo20213d} develop a three-dimensional (3D) DT framework of an intelligent transportation system (ITS) based on roadside sensing methods. However, it is impractical to anticipate that the sensing range of RSUs can totally cover the entire traffic \cite{barrachina2013road}. Consequently, a more realistic method is to enhance the utilization of vehicular onboard sensors, making CAVs also contribute to the modeling of DT, as discussed in \cite{tonguz2013cars, razdan2023polyverif, liu2023software}.

When it comes to the communications aspect, the concept of V2X provides a collaborative approach where multiple smart entities can cooperatively perceive the traffic environment \cite{zeng2022federated, he2023towards}. Hence, V2X has found a multitude of DT-like or DT-related applications within the field of autonomous driving and advanced driving assistant systems (ADAS). The authors in \cite{oubbati2020search} develop a software-defined network (SDN) to achieve global path planning for connected vehicles, based on vehicle-to-BS (V2BS) and vehicle-to-UAV (V2U) communications. In \cite{wagner2023spat}, authors establish a V2I network between road vehicles and traffic lights to build a traffic light DT and optimize traffic control. In our previous works \cite{li2023hetsdvn}, a software-defined vehicular network (SDVN) architecture for heterogeneous V2X is designed to realize cooperative perception and ensure safety in autonomous driving. Fundamentally, a DT represents a form of central cooperative perception, wherein sensory data is aggregated and fused at a central server. Thus, the establishment of a heterogeneous V2X network lays the groundwork for the extrapolation of the proposed SMDT platform.

In aligning with the discussed sensing and communication challenges, computation also emerges as a critical component in the DT frameworks, especially the collaboration of cloud and edge computing \cite{liu2022real}. Since edge computing is well-suited for handling delay-sensitive tasks and can also offload the cloud computation, it is often employed in some related research for functions like environmental perception and object detection \cite{tihanyi2021towards}, as well as autonomous driving system tasks like motion planning and motion control \cite{razdan2023polyverif}. On the other hand, the central cloud, as a data pool and high-level decision-maker, aggregates data from various smart entities and makes decisions, then provides feedback or services back to smart entities. For example, \cite{wang2022dtcoop} demonstrates the use of a first-in-first-out slot reservation algorithm within a centralized server for cooperative driving at non-signalized intersections. In another instance, \cite{dong2023mixed} presents a cloud control testbed for validating multi-vehicle cooperation and vehicle-road-cloud integration. Hence, in our platform, the synergetic integration of cloud and edge computing necessitates careful consideration.

\subsection{Contributions}

In Table \ref{tab: compare}, we summarize some related works and compare the main features of their research, where the level of driving automation proposed by the Society of Automotive Engineers (SAE) is applied to assess the degree of autonomy in these studies. To our knowledge, there are few studies investigating the confluence of DT technology and autonomous driving. Additionally, these studies predominantly employ validation methods of simulation or field trials. Simulation-based research not only establishes DT models but can also derive a series of new services. However, those conducting field trials face challenges in demonstrating the feedback from the digital to the physical world, where the concept of DT is conceived as a simulation environment for modeling real-world entities with high fidelity. In our perspective, a DT transcends mere mapping and reconstruction of the physical space within cyberspace. It should also enable an automatic bidirectional data flow between the physical and digital worlds to fully unlock the potential of DTs. i.e., leveraging global information in DT to optimize and control physical entities.

To validate the feasibility of applying the SMDT platform within the domain of autonomous driving and explore its potential, the most important task is the system design and implementation in the real world, which poses numerous challenges. Firstly, it is essential to create a real-time virtual representation of real-world traffic environments (i.e., traffic DT), which involves sensing, perceiving, and visualizing various static and dynamic entities. Secondly, the system design should consider safety and robustness issues, because of massive data collection from traffic environments and the requirement for autonomous vehicles to operate in unpredictable conditions, which necessitates an effective solution for dynamic traffic information acquisition and a reliable V2X communication network. Thirdly, successful implementation requires a holistic consideration and integration of hardware, software, and communication, as well as collaboration between cloud and edge computing. Finally, we also need to design effective methods for functionality verification and performance assessment, which involves substantiating functional capabilities through proof-of-concept (PoC) experiments and appraising the benefits brought by SMDT platform via large-scale traffic simulations.

Hence, based on the discussions above, the main contribution of this paper is a novel SMDT platform that can provide cloud services for CAVs. In particular, we make the following key contributions:
\begin{itemize}
  \item \emph{Architecture design and use case}: we develop a novel system architecture of the SMDT platform designed to address safety and robustness considerations. This architecture integrates cloud and edge computing across RSUs, CAVs, and a central cloud. Then we introduce a CAV navigation system as the use case of the SMDT platform;
  \item \emph{Comprehensive system design and implementation}: we detail the intricacies of core software and hardware components, as well as pivotal technologies for realizing SMDT platform in real-world autonomous driving environment. Then we achieve the implementation in the test field, including a cloud server and distributed intelligent RSU and CAV edges with sensor, communication module, and edge computing capability;
  \item \emph{PoC demonstration}: we conduct a field trial on SMDT-empowered route planning service to validate the system functionalities. We also evaluate the communication reliability and latency based on standards for SSMS and information sharing V2X use cases proposed by the 3rd generation partnership project (3GPP);
  \item \emph{Large-scale traffic simulation}: we develop a traffic simulation solution so as to study the benefits of the proposed SMDT platform in terms of traffic efficiency and safety when using the proposed CAV navigation system.
\end{itemize}

The rest of this paper is organized as follows. Section II explains the system architecture of SMDT platform and introduces the CAV navigation system as a use case, then we analyze the communication and latency requirements for the proposed SMDT platform and CAV navigation system. Then, the deployment intricacies, including software installation, hardware deployment, and communication establishment, are presented in Section III. Section IV demonstrates the case study in the PoC experiment to validate the system functionality. In Section V, the results from large-scale traffic simulation are discussed, and some metrics are measured to validate the system efficacy from the perspective of reliability and latency. Finally, we draw conclusions in Section VI.

\section{SMDT Platform for Autonomous Driving}
\begin{figure*}[!t]
    \centerline{\includegraphics[width=0.8\textwidth]{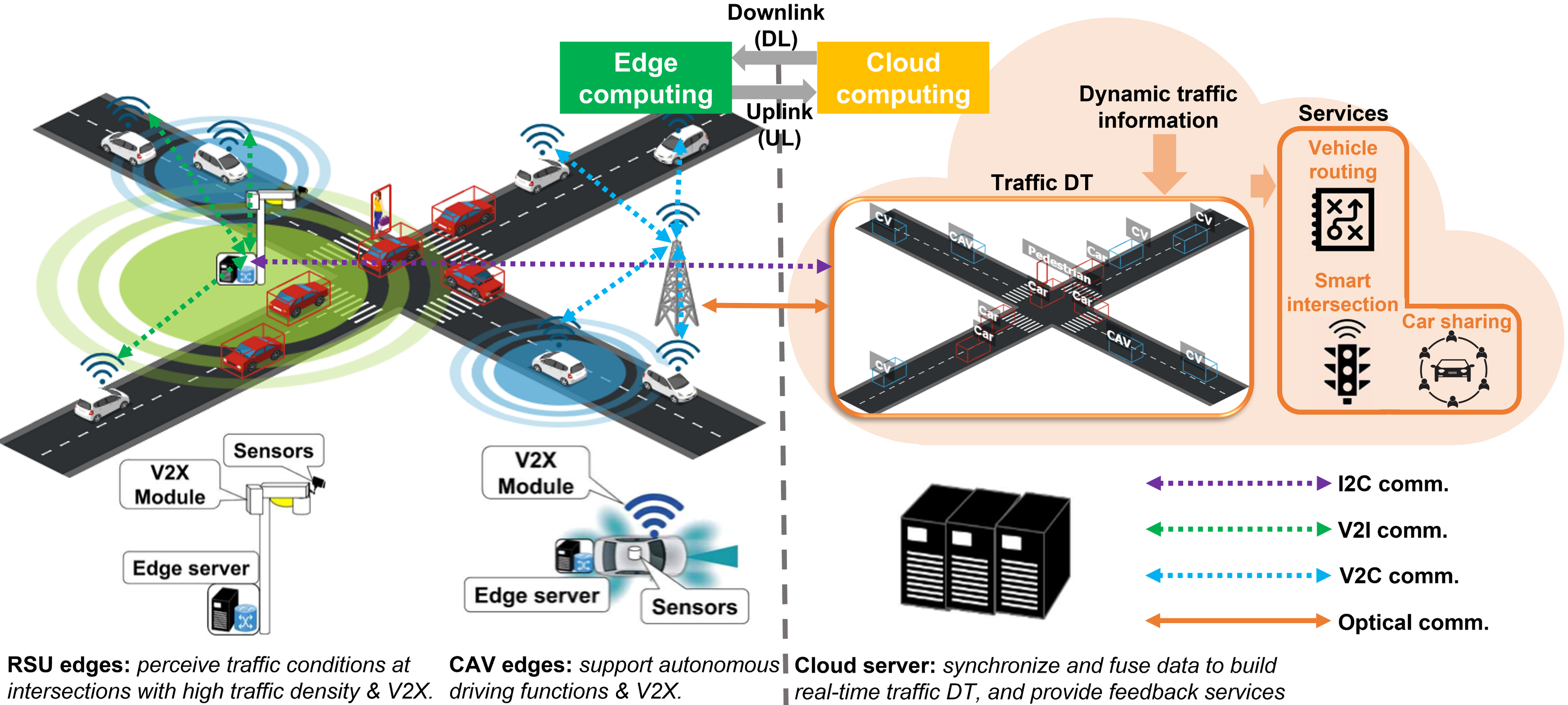}}
    \caption{SMDT high-level conceptual system architecture with cloud/edge computing.}
    \label{fig: cloudedge}
\end{figure*}

In this section, we introduce the architecture of the SMDT platform, in which a \emph{traffic DT} replicates the real-time real-world traffic information in cyberspace and enables enhanced cloud services for traffic efficiency and road safety. Relying on the potential of the SMDT platform, we propose the DT-empowered CAV navigation system, of which the functionality and latency requirements are discussed in detail.

\subsection{SMDT Platform Architecture}

The proposed SMDT platform integrates the cloud and edge computing resources of ITS. This architecture is chosen to exploit the robust computational capabilities of the cloud for data-intensive tasks while utilizing edge computing to minimize response times for real-time services. This dual-computing approach is essential to manage latency, safety, and processing efficiency in large-scale traffic network operations effectively. Computing at the cloud can utilize plentiful and stronger processing units, but it introduces considerable network latency. The SMDT platform prioritizes cloud computing when handling computation-intensive tasks that require global big data from large-scale traffic networks. Computing at the edge cannot enjoy the same processing efficiency as cloud computing due to the limited power supply and space for deploying units. However, it guarantees real-time responses to the users, addressing their demands for delay-sensitive services on the SMDT platform. As depicted in Fig.~\ref{fig: cloudedge}, the computing units of the SMDT platform are categorized into three types: RSU edges, CAV edges, and a central cloud. They are interconnected and collaboratively support the platform operations, including \emph{traffic DT} creation, data processing, and user interaction.

\emph{1) RSU Edges:} RSUs play an important role in the SMDT platform, primarily focused on high-traffic areas. RSUs are equipped with a variety of sensors to monitor traffic flow and detect accidents, and they also have the capability to receive data from connected vehicles (CVs) through V2I communication, which allows for a more efficient and targeted deployment of RSUs, ensuring they are strategically placed where their impact on traffic management and safety is maximized. These two roles of RSU enhance the effectiveness of the \emph{traffic DT} and optimize the utilization of RSUs in the overall network.

\emph{2) CAV edges:} All CAVs operate in a highly automated mode and strictly adhere to directives issued by the cloud. Besides contextual awareness, CAV sensors are used for localization, local path planning, and motion control. As a basic requirement of the SMDT platform, CAVs should report their positions to get services and upload their sensing data to compensate for the undetectable areas of RSU sensors when constructing the \emph{traffic DT}. In essence, CAVs are the mobile counterparts of RSUs. Importantly, these CAVs should be designed with a monitoring system that can be aware of the signal loss and switch to standalone mode for decision-making, ensuring that even if a CAV temporarily loses its communication link, it can still maintain safety and operational integrity until the connection is re-established.

\emph{3) Central cloud:} The cloud is the place where a dynamic \emph{traffic DT} is established by integrating a comprehensive array of traffic data. This integration encompasses real-time information from both RSU and CAV edges, along with inputs from various commercial traffic information providers. Serving as a vital resource pool, the cloud processes and analyzes this diverse data to offer a detailed and expansive view of traffic conditions. Such a robust integration of multiple data sources is instrumental for both providing accurate traffic analysis and ensuring the robustness of the system, thereby enhancing management solutions for road safety and traffic efficiency.

Data sharing between cloud and edge computing can be divided into downlink (DL) and uplink (UL). The edge servers at the RSU and CAV continuously upload various levels of real-time sensing data, such as raw data or processed data, according to different use cases. The cloud provides services and issues directives over the DL to edges. To ensure connectivity among these smart entities and reliable DL/UL between cloud and edge, it is necessary to establish a V2X network, including V2C, V2I, and infrastructure-to-cloud (I2C) communications, where V2C communication is designed with a multi-radio access technology (Multi-RAT) handover mechanism based on channel conditions to ensures consistent and stable communication. V2C communication ensures the reach to the cloud through cellular networks or relayed by RSUs using dedicated short-range communications (DSRC), millimeter-wave (mmWave), and Wi-Fi.

\subsection{Use Case: CAV Navigation System}

To evaluate and highlight the benefits of the proposed SMDT platform from the practical viewpoint of applications and services, a CAV navigation system is designed based on SMDT platform. In a dynamic traffic environment, road segments may experience unexpected incidents—such as accidents, large gatherings, or peak-time congestion—that can drastically affect journey time and road safety. Therefore, it is crucial to devise a mechanism that continuously detects such incidents and dynamically re-routes users.

Hence, the CAV navigation system is designed to employ an event-triggered planning strategy. As users enter the road network, the optimal routes are generated based on current traffic conditions to minimize journey time from origin to destination. During transit, if specific events occur within the traffic network, such as a car crash, a route re-planning service is triggered to help users avoid these events. 

\subsubsection{Journey Time Calculation}

We use a directed graph $G = (N, L)$ to represent the traffic network. $N$ is the set of nodes (i.e., traffic intersection) and $N=\{n^i\}^M_{i=1}$, where $M$ is the total number of nodes. $L$ is defined as the set of directed links (i.e., unidirectional traffic roads) and $L=\{l^{i,j}=(n^i \rightarrow n^j), n^i,n^j \in N\}$, where $l^{i,j}$ means the directed road from $n^i$ to $n^j$. As we can obtain the real-time traffic volume $x$ on the road link $l$ with the proposed SMDT platform, based on the traffic volume and road length $s$, we can calculate the traffic density $k$ with:

\begin{equation}
    k = \frac{x}{s}
\end{equation}

According to \cite{cheung2012dora, xing2015maximum}, the relationship between traffic density $k$ and journey speed $v$ can be modeled as

\begin{equation}
    v = v_{\textrm{free}}(1-\frac{k}{k_{\textrm{max}}})
\end{equation}

\noindent where $v_f$ is the free-flow speed, and $k_{\textrm{max}}$ is the maximum vehicle density. Thus, the journey time can be calculated as:

\begin{equation}
    t_{\textrm{jry}} = \frac{s}{v} = \frac{s}{v_{\textrm{free}}(1-\frac{k}{k_{\textrm{max}}})}
\end{equation}

Based on equation (3), we use a matrix $T_{\textrm{jry}}$ to represent the node-to-node journey time for each road segment within the traffic network:

\begin{equation}
\boldsymbol{T_{\textrm{jry}}} =
\begin{pNiceArray}{cccc}[first-row,first-col]
           & n^1         & n^2              & \cdots   & n^{M}      \\
  n^1      & t^{1,1}(t)  & t^{1,2}(t)       & \cdots   & t^{M,M}(t)    \\
  n^2      & t^{2,1}(t)  & t^{2,2}(t)       & \cdots   & t^{M,M}(t)    \\ 
  \vdots   & \vdots      & \vdots           & \ddots   & \vdots     \\
  n^{M}    & t^{M,1}(t)  & t^{M,2}(t)       & \cdots   & t^{M,M}(t)    \\
\end{pNiceArray}
\end{equation}

\noindent where $M$ is the total number of network nodes. If there exists no link from node $n^i$ to node $n^j$, then the journey time $t^{i,j}$ will be set to $+\infty$.

\subsubsection{Event-triggered Mechanism}

Traffic event is defined as unexpected and sporadic scenarios that have severe impacts on traffic efficiency and safety. In this study, we primarily analyze the two typical scenarios along with their detection criteria:

\begin{itemize}
\item \emph{Pedestrian gathering:} To detect the occurrence of overcrowded gatherings, we introduce a pedestrian density threshold $d_{\textrm{thre}}$ as the criteria for assessment. When the measured density $d^i$ in the vicinity of the intersection $n^i$ is higher than $d_{\textrm{thre}}$, the intersection $n^i$ will be regarded as overcrowded.

\item \emph{Traffic accidents:} 
Vehicular speeds can be regarded as a criterion of the accident occurrence. Specifically, if a particular link $l^{i,j}$ or intersection $n^i$ exhibits vehicles with speeds consistently below a threshold $v_{\textrm{thre}}$, it is inferred that the link $l^{i,j}$ or the intersection $n^i$ has become a locus of the traffic accident.
\end{itemize}

In the traffic DT-based traffic monitoring process, we continuously update the sets of nodes $N_{\textrm{eve}}$ and links $L_{\textrm{eve}}$ that are influenced by the aforementioned two kinds of events. 

\subsubsection{Workflow of CAV Navigation System} 

Based on the computation of \emph{journey time} and the \emph{event-triggered mechanism}, the workflow of CAV navigation system can be designed as a cooperative event-triggered route planning process. The cooperative aspect is manifested by the necessity for all CAV users to upload their sensor data. This collaborative data sharing serves to effectively compensate for areas not covered by RSU sensors. On the other hand, the \emph{event-triggered mechanism} comes into play when traffic events are detected within the network. Such incidents then trigger route re-planning for users whose original routes overlap with these events. 

\begin{algorithm}[t]
\SetAlgoLined
\KwIn{  \\
        $G=(N, L)$  // Road network \\
        $\boldsymbol{T_{\textrm{jry}}}$  // Journey time of each road link \\
        $U=\{u^i\}_{i=1}^{M_U}$  // Set of CAV user already in road network \\
        $U_{\textrm{new}}=\{u_n^i\}_{i=1}^{M_N}$  // Set of new CAV user entering road network \\
        $R(u_i)$ // Planned route of each user already in road network \\
        $N_{\textrm{eve}}$ // Set of intersections with detected event \\
        $L_{\textrm{eve}}$ // Set of roads with detected event}
\KwOut{ \\
        $R_{\textrm{new}}(u_i)$ // Re-planned route for each user \\}
 Initialization: $R_{\textrm{new}} \leftarrow \emptyset$\;
// Plan route for new entering CAVs. \\
\For{$u_n^i \in U_{\textrm{new}}$}{$n_{\textrm{start}} := u_n^i.CurrentPosition$\; \\
                          $n_{\textrm{end}} := u_n^i.Destination$\; \\
                          $R_{\textrm{new}}(u_n^i) := \textrm{Dijkstra}(N, L, \boldsymbol{T_{\textrm{jry}}}, n_{\textrm{start}}, n_{\textrm{end}})$\;
                          }
// Set the journey time of links pointing to the nodes, where the events occurred, to infinity. \\
\For{$n^i \in N_{\textrm{eve}}$}{\For{$n^j \in N_{\textrm{eve}}\setminus \{n^i\}$}{\If{$\boldsymbol{T_{\textrm{jry}}}(n^j,n^i) \neq +\infty$}{$\boldsymbol{T_{\textrm{jry}}}(n^j,n^i) := +\infty$}
                                                                }
                       }
// Set the journey time of links with events to infinity. \\
\For{$l^{i,j} \in L_{\textrm{eve}}$}{$\boldsymbol{T_{\textrm{jry}}}(n^i,n^j) := +\infty$}
// Re-plan route for users. \\
\For{$u^i \in U$}{\For{$l^{i,j} \in R(u_i)$}{\If{there exists $\boldsymbol{T_{\textrm{jry}}}(n^j,n^i) = +\infty$}{$n_{\textrm{start}} := u^i.CurrentPosition$\; \\
                                                                                            $n_{\textrm{end}} := u^i.Destination$\; \\
                                                                                            $R_{\textrm{new}}(u^i) := \textrm{Dijkstra}(N, L, \boldsymbol{T_{\textrm{jry}}}, n_{\textrm{start}}, n_{\textrm{end}})$}
                                            }
                 }
\caption{Cooperative Event-triggered Route Planning}
\end{algorithm}

The overall operation of the proposed route planning algorithm is shown in Algorithm 1. The procedure operates on the road network $G$ and the predicted journey \textbf{$T_{\textrm{jry}}$}. For the CAV users just entering the network, denoted as $U_{\textrm{new}}$, their initial route is generated using Dijkstra algorithm. Here, we note that the Dijkstra algorithm is employed to generate the fastest path, which is facilitated by assigning each road link's journey time as its weight. If some events are detected at intersections, the journey time for links pointing towards these intersections will be set to infinity, to help users circumvent such events. Similarly, if events happen on specific road links, their journey time will also be set to infinity. Then the planned routes of CAV users should be re-evaluated. If any part of a user's planned route includes a link with an infinite journey time, the user's route is recalculated using Dijkstra algorithm, which ensures that all users are always provided with an optimal route considering the latest information.

\subsection{Requirements for SMDT-based CAV Navigation Use Case}

\begin{figure}[t]
    \centerline{\includegraphics[width=0.48\textwidth]{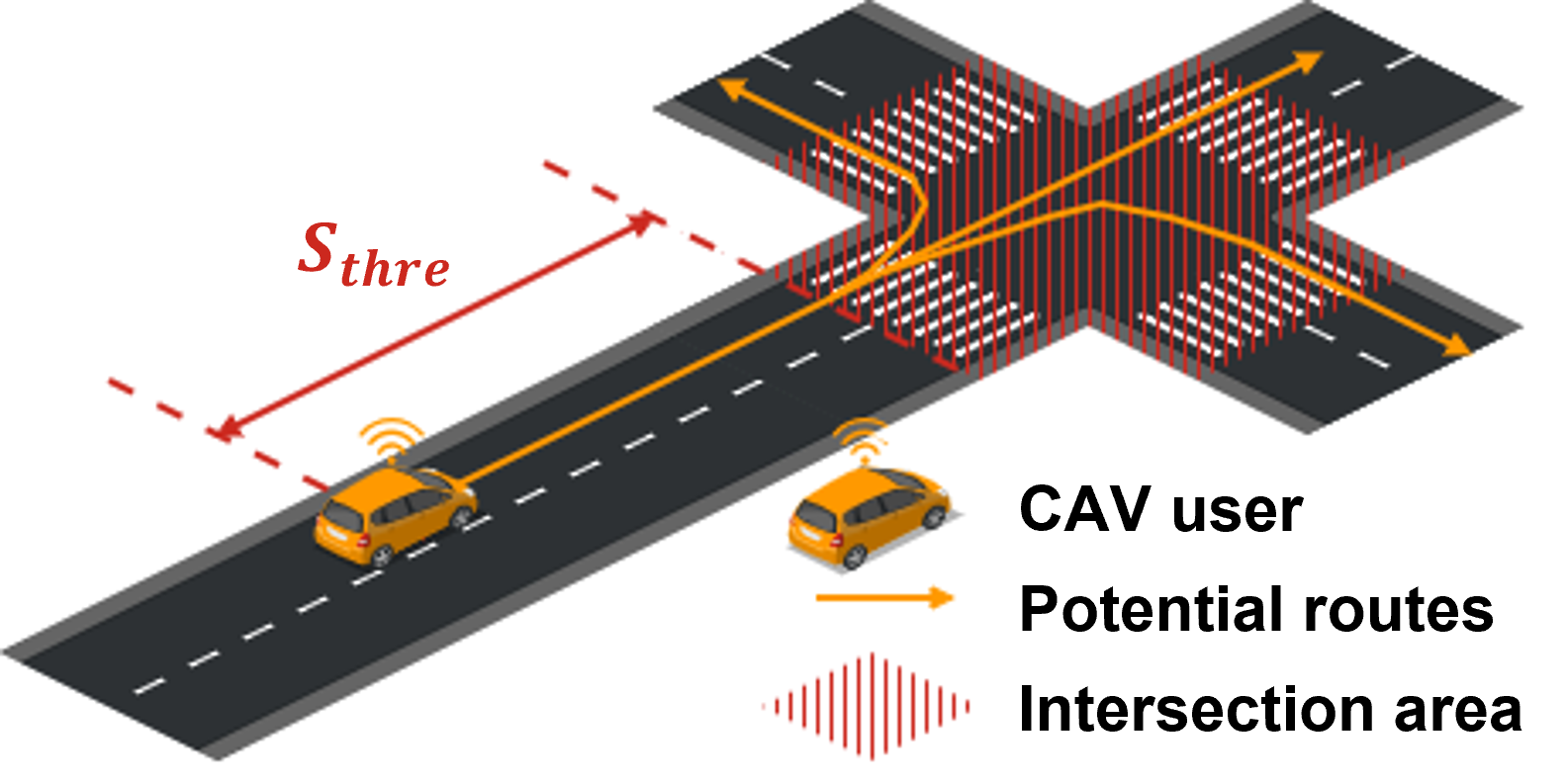}}
    \caption{Illustration of route-planning request distance.}
    \label{fig: req}
\end{figure}

The SMDT-based CAV navigation system features a centralized architecture focusing on two core processes: \emph{traffic DT modeling} and \emph{route planning service}. To achieve \emph{traffic DT modeling} over the cloud plane with RSU and CAV sensing data, sensor-equipped RSUs are adopted for collecting raw data nearby traffic intersections, and CAVs with onboard sensors driving within the road network serve to supplement traffic conditions beyond the sensing range of RSUs. As for \emph{route planning service}, it is important to establish reliable communication to ensure the dissemination of planned routes. Upon near entering into the road network, the user should ``tell" the cloud its impending presence, initiating continuous uploads of its position and desired destination, and then the cloud responds by furnishing an initially planned route. Since the initial planned route might make the user alter their predetermined direction at an intersection, as shown in Fig.~\ref{fig: req}, the request should be initiated prior to the user entering the intersection. Here, we define a threshold distance $S_{\textrm{thre}}$ at intersections, i.e., when the distance between the CAV user and the intersection area reaches $S_{\textrm{thre}}$, the user will send a route planning request to the cloud.

The threshold distance $S_{\textrm{thre}}$ is determined with the consideration of comfort issues. The vehicle should have enough time to process the planned route before reaching the intersection. Given the presence of pedestrians crossing at intersections, it is necessary for the vehicle to comfortably decelerate and stop before reaching the intersection. According to the Institute of Transportation Engineers (ITE) recommendations, a comfortable deceleration rate $a_{\textrm{comfy}}$ should be less than $10$ ft/s\textsuperscript{2}, equivalent to $3.048$~m/s\textsuperscript{2} \cite{ite1985determining}. Hence, $S_{\textrm{thre}}$ can be decided with the maximum speed (i.e., free-flow speed $v_{\textrm{free}}$) as:

\begin{equation}
    S_{\textrm{thre}} = \frac{v_{\textrm{free}}^2}{2 a_{\textrm{comfy}}} = 0.164~v_{\textrm{free}}^2
\end{equation}

\begin{table}[t]
\centering
\caption{V2X Use Cases and Respective KPIs}
\label{tab:v2x}
\begin{tabular}{c|ccc}
\hline
Use cases                                                     & \begin{tabular}[c]{@{}c@{}}Reliability\\ (\%)\end{tabular} & \begin{tabular}[c]{@{}c@{}}E2E latency\\ (ms)\end{tabular} & Total latency                       \\ \hline
\multirow{2}{*}{SSMS}                                         & \multirow{2}{*}{95}                                        & \multirow{2}{*}{10}                                        & \multirow{2}{*}{$T_{\textrm{dt}}\ll T_{\textrm{svc}}$} \\
                                                              &                                                            &                                                            &                                     \\
\begin{tabular}[c]{@{}c@{}}Information\\ Sharing\end{tabular} & -                                                        & 100                                                        & $T_{\textrm{svc}}<0.164~v_{\textrm{free}}$                 \\ \hline
\end{tabular}
\end{table}

Communication links in the \emph{traffic DT modeling} and \emph{route planning service} processes correspond to two V2X use cases proposed by 3GPP \cite{garcia2021tutorial}, i.e., SSMS and information sharing for high/full automated diving, respectively. SSMS enables the sharing of raw or processed sensor data to build collective situational awareness, which allows the central server to access and gather real-time traffic information obtained by RSU sensors. The information sharing use case encompasses two scenarios: cooperative perception and cooperative maneuver. Cooperative maneuver necessitates the sharing of detailed planned trajectories among all participating vehicles via V2X communication, which permits the central server to exert control over the vehicular routes.

To assess whether the proposed SMDT platform can meet the necessary requirements of the above two use cases, the respective KPIs are summarized in Table \ref{tab:v2x}. The requirements in terms of reliability and E2E latency are defined in the same way as 3GPP technical report (TR) 22.886 \cite{3gpp}. By definition, reliability refers to the probability of successful transmission reliability and E2E latency refers to the one-way time delay it takes to deliver a piece of information from a source to a destination, without the application-layer processing delay. In this study, E2E latency of SSMS and information sharing use cases are defined respectively as the unidirectional communication delay between RSUs and the cloud, termed as $T_{\textrm{I2C}}$, and between the CAVs and the cloud, termed as $T_{\textrm{V2C}}$. In addition, we also define the total latency of \emph{traffic DT modeling} and \emph{route planning} processes. In our system, the basis for decision-making is the dynamic traffic DT, which must represent real-time traffic conditions. Consequently, the latency in DT modeling $T_{\textrm{dt}}$ needs to be significantly shorter than the latency in the route planning service process $T_{\textrm{svc}}$, to ensure that the traffic DT used for decision-making can be regarded as real-time. The DT modeling latency $T_{\textrm{dt}}$ mainly consists of RSU edge computational time and one-way E2E latency from LiDAR to cloud server, expressed as:

\begin{equation}
    T_{\textrm{dt}} = T_{\textrm{RSU}} + T_{\textrm{I2C}}
\end{equation}

Considering the route planning service is based on a request-response mechanism, it is imperative to ensure that users can receive, load, and execute the new planned route before entering the intersection. The route planning latency $T_{\textrm{svc}}$, spanning from the moment the user reaches the request distance $S_{\textrm{thre}}$ to the execution of the route, should meet:

\begin{figure*}[!t]
    \centerline{\includegraphics[width=0.85\textwidth]{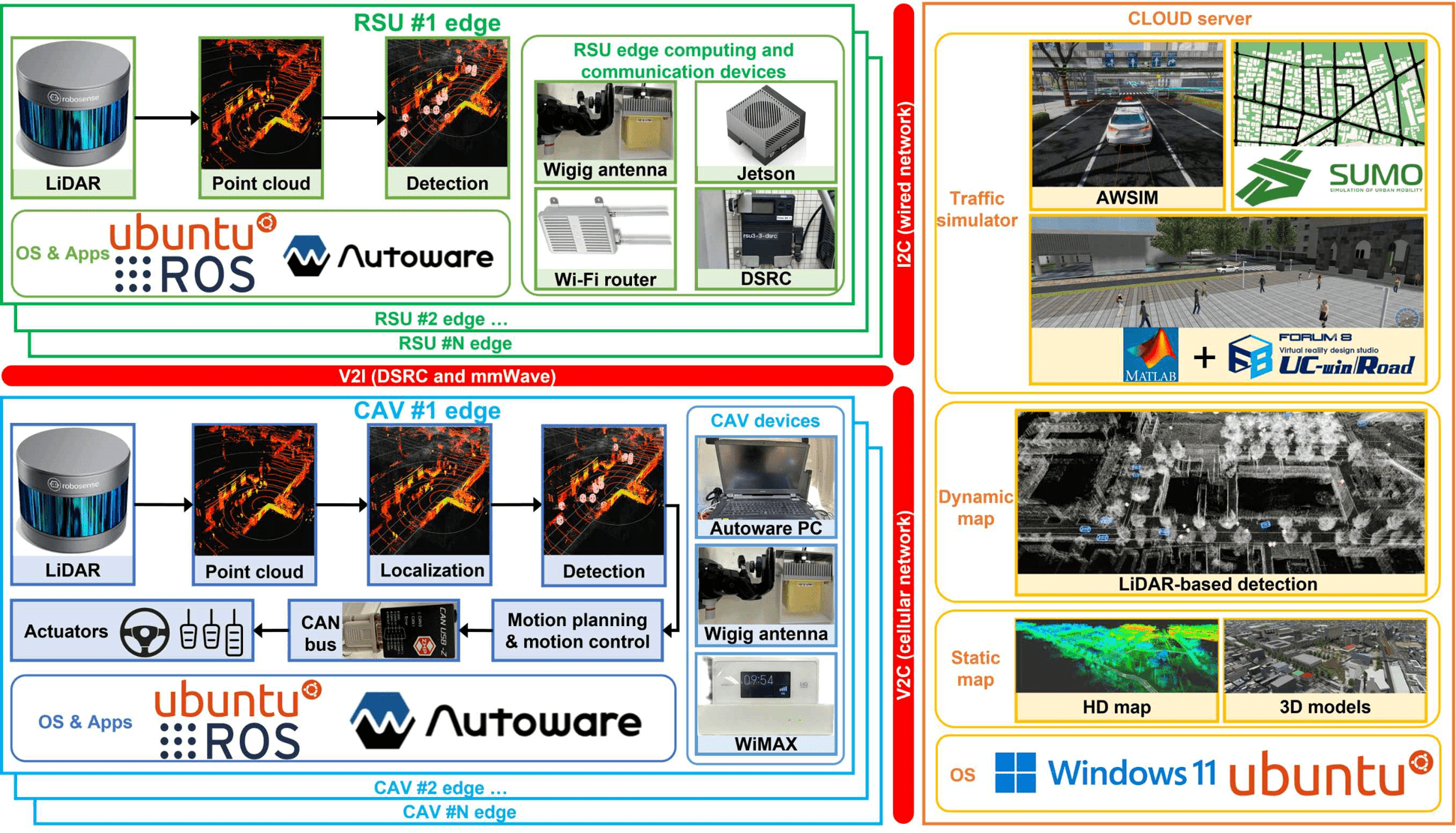}}
    \caption{Example system design of the SMDT platform. }
    \label{fig: dt-design}
\end{figure*}

\begin{equation}
    T_{\textrm{svc}} \leq \frac{S_{\textrm{thre}}}{v_{\textrm{free}}} = 0.164~v_{\textrm{free}}
\end{equation}

\noindent where the latency of offering route planning service $T_{\textrm{svc}}$ includes time delays caused by localization $T_{\textrm{local}}$, route loading and execution $T_{\textrm{exe}}$ in the autonomous driving system, cloud computing $T_{\textrm{cloud}}$, and bidirectional V2C communication $T_{\textrm{V2C}}$. 

\begin{equation}
    T_{\textrm{svc}} = T_{\textrm{local}} + T_{\textrm{exe}} + T_{\textrm{cloud}} + 2~T_{\textrm{V2C}}
\end{equation}

\section{Design and Implementation of SMDT Platform}

\begin{figure*}[ht]
    \centerline{\includegraphics[width=0.85\textwidth]{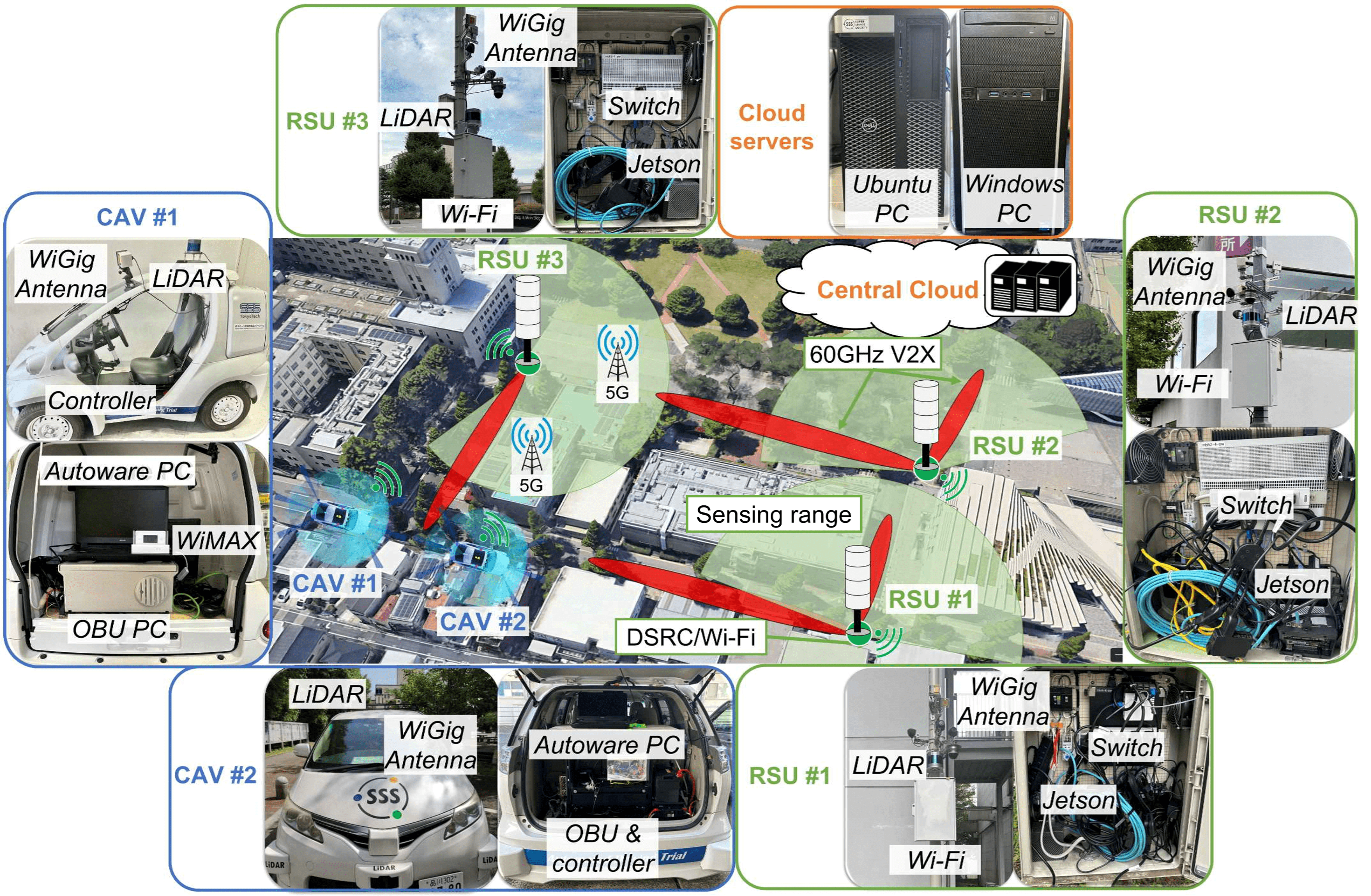}}
    \caption{Hardware components in smart mobility R\&E field of Tokyo Tech Academy for Super Smart Society.}
    \label{fig: implementation}
\end{figure*}

\begin{table*}[t]
\centering
\caption{Communication Requirements and Performances}
\label{tab:commu}
\begin{tabular}{c|cccc}
\hline
Function             & Requirements                                                                                      & Message contents                                                                                                                       & Methods      & Performances                                                                       \\ \hline
V2C                  & \begin{tabular}[c]{@{}c@{}}Data rate: $\geq 80$~Mbps\\ Coverage: $\geq 1$~km\end{tabular} & \begin{tabular}[c]{@{}c@{}}Uplink: vehicle position and processed sensor data. \\ Downlink: cloud-based decisions\end{tabular}        & WiMAX        & \begin{tabular}[c]{@{}c@{}}$\geq 120$~Mbps\\ $\geq 50$~km\end{tabular} \\ \hline
\multirow{2}{*}{V2I} & \begin{tabular}[c]{@{}c@{}}Data rate: $\geq 1$~Mbps\\ Coverage: $\geq 50$~m\end{tabular}  & \begin{tabular}[c]{@{}c@{}}Cooperative perception (processed data, \\ e.g., detection and tracking result)\end{tabular}       & Wi-Fi router & \begin{tabular}[c]{@{}c@{}}$\geq 10$~Mbps\\ $\geq 200$~m\end{tabular}   \\ \cline{2-5} 
                     & \begin{tabular}[c]{@{}c@{}}Data rate: $\geq 1$~Gbps\\ Coverage: $\geq 20$~m\end{tabular}  & \begin{tabular}[c]{@{}c@{}}Cooperative perception (raw sensor data, e.g. \\ LiDAR point cloud and camera images)\end{tabular} & WiGig        & \begin{tabular}[c]{@{}c@{}}$\geq 1$~Gbps\\ $\geq 120$~m\end{tabular}    \\ \hline
I2C                  & Data rate: $\geq 1$~Gbps                                                                        & Environmental perception (both raw data and processed data)                                                                   & Ethernet     & $\geq 1$~Gbps                                                                \\ \hline
\end{tabular}
\end{table*}

\begin{table*}[t]
\centering
\caption{Hardware Devices and Specifications}
\label{tab:hard}
\begin{tabular}{c|ccc}
\hline
\multirow{2}{*}{Type}                                                             & \multicolumn{3}{c}{Description}                                                                                        \\ \cline{2-4} 
                                                                                  & Device names     & Specifications                                                                              & Amount \\ \hline
\multirow{2}{*}{CAV}                                                              & RoboCar (CAV \#1)         & Automated vehicle with driving controller                                                   & 1      \\
                                                                                  & MiniVan (CAV \#2)        & Automated vehicle with driving controller                                                   & 1      \\ \hline
\multirow{2}{*}{Sensor}                                                           & RS-LiDAR-32     & Location: CAV edges, Range: 200 m, Accuracy: $\pm$ 3 cm, Rotation Speed:10/20 Hz            & 2      \\
                                                                                  & RS-LiDAR-80     & Location: RSU edges, Range: 200 m, Accuracy: $\pm$ 3 cm, Rotation speed: 5/10/20 Hz         & 3      \\ \hline
\multirow{5}{*}{\begin{tabular}[c]{@{}c@{}}Cloud/\\ edge \\ servers\end{tabular}} & Autoware PC     & Location: CAV edge (RoboCar), OS: Ubuntu 16.04, ROS: Kinetic, Autoware. AI                  & 1      \\
                                                                                  & Autoware PC     & Location: CAV edge (MiniVan), OS: Ubuntu 22.04, ROS: Foxy, Autoware. Universe               & 1      \\
                                                                                  & Jetson AGX Orin & Location: RSU edges, OS: Ubuntu 20.04 (JetPack 5.0.2), ROS: Galactic, Autoware. Universe    & 3      \\
                                                                                  & DT engine       & Location: cloud server, OS: Ubuntu 20.04, ROS: Galactic, Autoware. Universe & 1      \\
                                                                                  & DT simulator    & Location: traffic simulator, OS: Windows 11, Simulation app: SUMO, UC-win/Road              & 1      \\ \hline
\end{tabular}

\end{table*}

In this section, we design an example system architecture of the proposed SMDT platform tailored for implementation, as shown in Fig.~\ref{fig: dt-design}. In the vein of conceptual design, the example design also includes three key components: RSU, CAV, and the central cloud. The primary objective of this implementation-tailored system is to realize the conception of SMDT platform depicted in Fig.~\ref{fig: cloudedge}, utilizing LiDAR sensing, cloud-edge computing, and V2X communication, which ensures that the \emph{traffic DT modeling} and \emph{route planning service} can be developed and deployed in the real-world traffic environment.

\subsection{Hardware Deployment and Heterogeneous V2X Network}

\begin{figure}[t]
\centering
\begin{minipage}[b]{0.45\textwidth}
\includegraphics[width=1\textwidth]{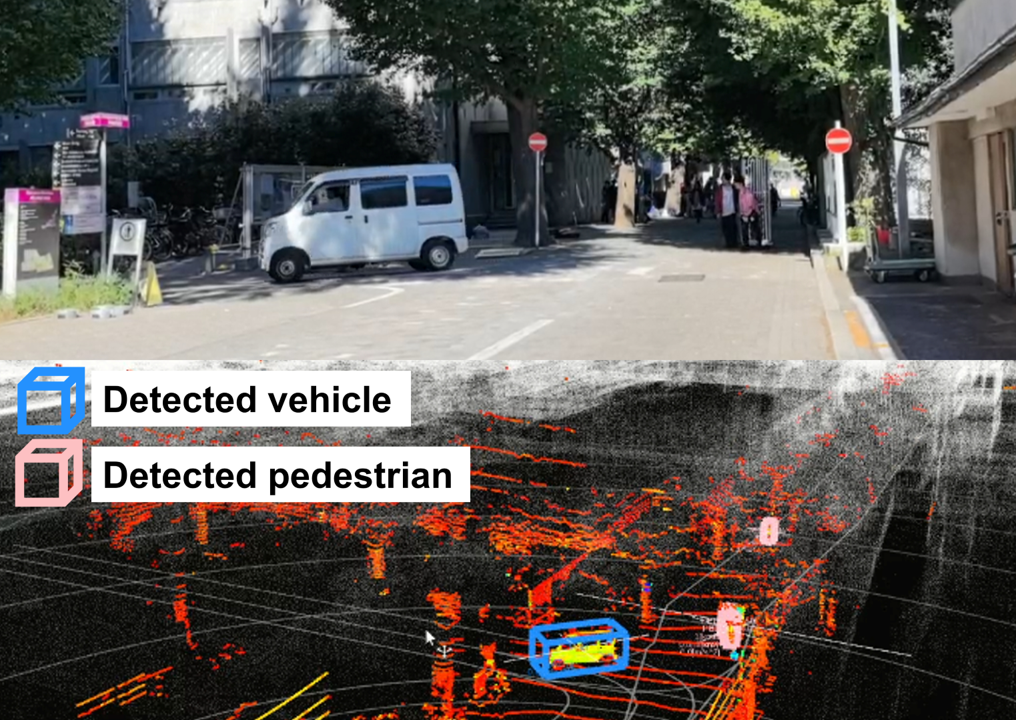} 
\end{minipage}
\caption{An example of LiDAR-based object detection in the world.}
\label{fig: detection}
\end{figure}

Our SMDT testing environment is named Tokyo Tech Smart Mobility R\&E Field\footnote{https://www.wise-sss.titech.ac.jp/en/.}, as shown in Fig.~\ref{fig: implementation}. In the field setting, there are three RSUs installed at three corners of a square road section, two CAVs that can achieve high autonomy, as well as the remote cloud server located far away from the testing field. The communication system within the SMDT platform is based on a heterogeneous V2X network and managed by an SDN controller \cite{li2023hetsdvn}, which enables three kinds of V2X links: V2I, V2C, and I2C communications. Considering the large communication distance between vehicles and the cloud, the cellular network is applied for V2C communication. Depending on the application scenario, RSUs may send data of different levels to the cloud server and nearby vehicles, such as raw data or processed data, which have distinct requirements for communication coverage and bandwidth. Hence, we use wired networks for I2C communication and employ both DSRC and mmWave technologies for V2I. Hence, each RSU and CAV are equipped with communication modules, including WiMAX for V2C communication, a Wi-Fi router as a replacement for DSRC, WiGig antenna for mmWave communication, and Ethernet cables for wired communication through the campus network. In Table \ref{tab:commu}, we list the required communication speeds and coverage in our experiment field, as well as the performance of selected communication methods.

As for sensing and computing devices, summarized in Table \ref{tab:hard}, an 80-layer LiDAR is integrated alongside an NVIDIA Jetson in each RSU. LiDAR is responsible for capturing raw data (i.e., dynamic point clouds) from the physical world. The Jetson, on the other hand, is employed to implement specific functional modules, such as object detection and tracking based on point clouds, thereby enabling object-level projection from the physical world to the digital world. Two CAVs are outfitted with 32-layer LiDAR sensors positioned on the rooftop to sense their surroundings. A dedicated Autoware PC is utilized for processing the LiDAR data, facilitating environmental perception, localization, motion planning, and motion control. The control signals are then transmitted to the onboard unit (OBU) to achieve autonomous driving. The remote cloud server works as the central hub for data aggregation and storage, global information processing, and providing feedback services to the autonomous vehicle. Therefore, the performance requirements for the cloud server are very high, including scalability, processing power, and reliability to efficiently and securely handle large volumes of real-time data. Consequently, we select a computer equipped with a GeForce RTX 5000 GPU and ensure abundant memory and storage resources to satisfy the computational demands.

\subsection{Software Installation and Traffic DT Modeling}

The software in our platform is centered around Autoware \cite{autoware2022autoware} and Robot Operating System (ROS). Autoware facilitates the detection and tracking of traffic participants within the RSU sensor range. The detection module applies the CenterPoint framework \cite{yin2021center}, which can detect, identify, and visualize 3D objects from the LiDAR point clouds in real time. Then, a Multi-object Tracker module \cite{autoware2022tracker}, is responsible for assigning the detected objects with IDs and estimating their velocities. An example of detection and tracking results is shown in Fig.~\ref{fig: detection}. Additionally, a Normal Distributions Transform (NDT) algorithm is utilized for the LiDAR scan matching with the 3D point cloud map, which enables real-time localization with centimeter-level accuracy. The driving route is represented in the form of Waypoints, which includes a set of points with coordinates and desired speed. To perform local motion planning and motion control, a velocity planner adjusts the velocity based on the Waypoints to decelerate or accelerate in response to nearby objects and road characteristics, such as stop lines and traffic lights. Finally, a Pure Pursuit algorithm is employed to generate coordinated velocities and steering angles and follow the target Waypoints.

Over the cloud plane, since ROS facilitates seamless communication among distributed computers within a Local Area Network (LAN), the cloud server can easily access all ROS2-defined messages from RSU edges. Besides, the Hypertext Transfer Protocol (HTTP) communication is utilized to establish the V2C link, enabling the real-time detection results (a kind of ROS2-defined massages) uploading and planned route downloading on CAV edges. After acquiring detection results from RSUs and CAVs, we can fuse them and build the real-time object-level \emph{traffic DT} in Autoware visualization tool Rviz. Then, the cloud server utilizes objects' positions to align them with the corresponding road segments, which facilitates traffic monitoring on each road section, thereby realizing \emph{journey time calculation} and \emph{event-triggered mchanism} mentioned in Section II.B. Additionally, for large-scale traffic experiments that cannot be executed within our testing field, we also create the simulation environment with some traffic simulation software, such as SUMO\cite{behrisch2011sumo} and UC-win/Road\cite{ucwin}.




\section{Proof-of-Concept Demonstration of CAV Navigation System}
In this section, we present an outdoor PoC experiment of SMDT-based route planning for autonomous driving. The experimental setup for route planning process in the Tokyo Tech Smart Mobility Field is first explained. Then we study two practical cases that effectively reflect key functions of proposed system.

\subsection{PoC Setup in Tokyo Tech Smart Mobility Field}

\begin{figure}[t]
    \centerline{\includegraphics[width=0.48\textwidth]{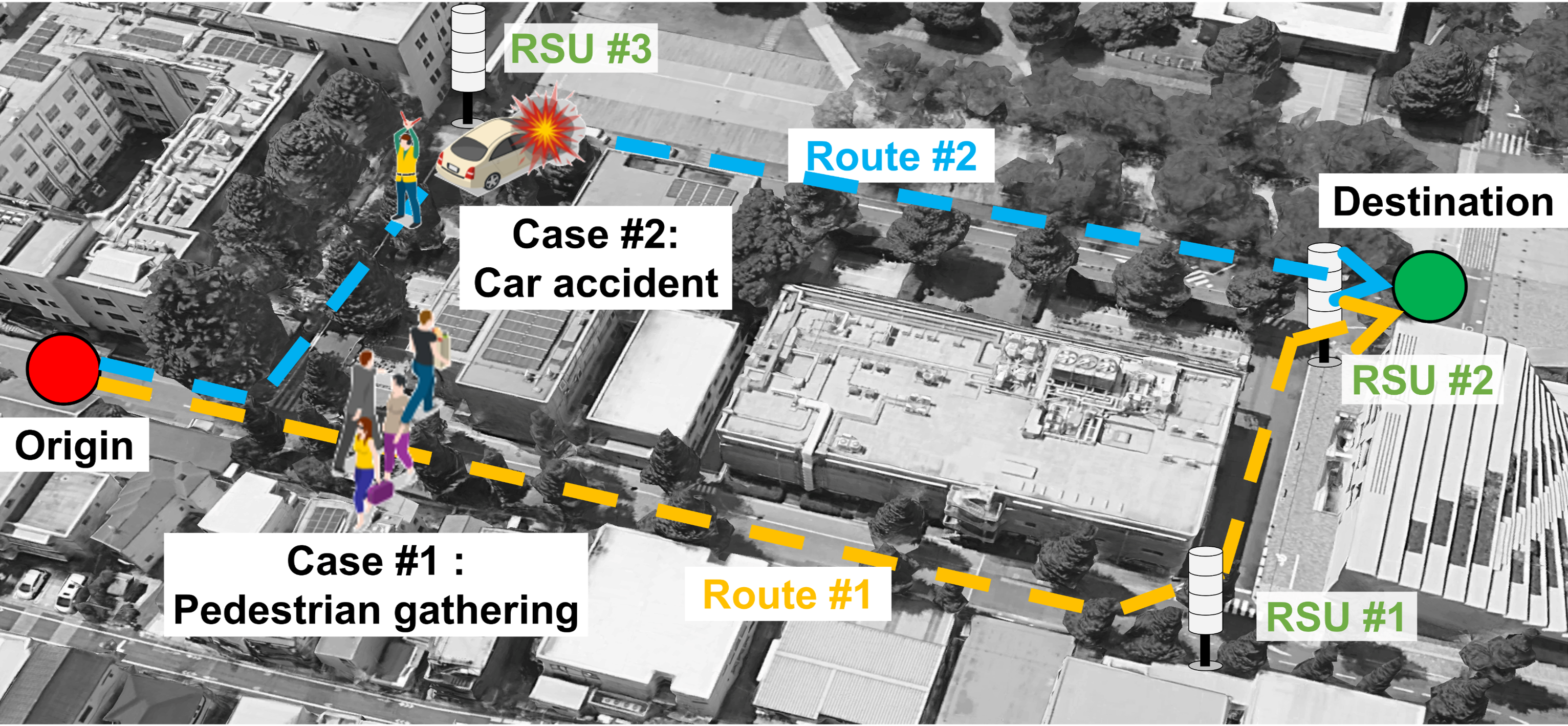}}
    \caption{An overview of experiment setup.}
    \label{fig: set}
\end{figure}

\begin{table}[t]
\centering
\caption{Roles of CAVs, RSUs, and the Cloud in PoC Experiment.}
\begin{tabular}{c|l}
\hline
Component                                              & \multicolumn{1}{c}{Functionalities}                                                                                                                                                           \\ \hline
RSU                                                    & \begin{tabular}[c]{@{}l@{}}1. Detect and track objects within sensing range;\\ 2. Detect traffic accidents within sensing range.\end{tabular}                                                  \\ \hline
CAV                                                    & \begin{tabular}[c]{@{}l@{}}1. Operate in a fully autonomous driving mode;\\ 2. Detect and track objects within sensing range;\\ 3. Detect traffic accidents within sensing range.\end{tabular} \\ \hline
\begin{tabular}[c]{@{}c@{}}Cloud\\ server\end{tabular} & \begin{tabular}[c]{@{}l@{}}1. Collect traffic information from RSUs and CAVs;\\ 2. Model and visualize DT;\\ 3. Assign routes for CAVs to avoid accidents.\end{tabular}                               \\ \hline
\end{tabular}
\label{table: roles}
\end{table}

Given the context of our Smart Mobility Field located within a campus environment, it inherently experiences minimal vehicular flow. This means that under normal conditions, vehicle volume is not a determinant of journey times. However, the driving environment is relatively complex due to the frequent presence of pedestrians and cyclists on the roadways. Coupled with a simple road network structure, our PoC experiment does not emphasize the \emph{journey time calculation} function or the usage of the Dijkstra algorithm to search for a route for autonomous vehicles. Instead, the focus is on demonstrating the \emph{event-triggered mechanism}, i.e., utilizing the established \emph{traffic DT} to detect and identify traffic events, then subsequently providing route planning for the CAV to bypass such incidents.

In this experiment, the ego vehicle corresponds to CAV \#1 as shown in Fig.~\ref{fig: implementation}. Its origin and destination are depicted in Fig.~\ref{fig: set}. There are two routes with approximately equal distances bridging the start and end points: Route \#1 ($206$~m) and Route \#2 ($210$~m), where Route \#1 is regarded as a default route. Under normal conditions without any traffic events, the cloud server would select Route \#1 and send its Waypoint file to CAV users by default. In Case \#1, A \emph{pedestrian gathering} event occurs on the default Route \#1, which is not within the sensing range of any RSU. Both CAV \#1 (i.e., ego vehicle) and CAV \#2 drive through the road network sequentially from origin to destination, where the maximum speed of CAVs $v_{\textrm{max}}$ are set to match the road speed limit, i.e., $20$~km/s in the campus. As CAV \#2 first enters the network and travels along the default Route \#1, it will encounter the \emph{pedestrian gathering} and stop in front of it at a safe distance. This event will be thus detected by CAV \#2 sensor. In Case \#2, we park CAV \#2 on Route \#2 to simulate the occurrence of a \emph{traffic accident}, positioned within the sensing range of RSU \#3. Then this event can be detected by RSU \#3. Some main responsibilities of RSUs, CAVs, and the cloud server in the PoC experiment are summarized in Table V.

\subsection{Case Study: Traffic DT-based Route Planning}

According to these two cases, a PoC experiment is conducted and a demonstration video\footnote{https://youtu.be/3waQwlaHQkk} is available online, showcasing the practical implementation and effectiveness of the SMDT-based CAV navigation system in real-world scenarios.

\subsubsection{Case \#1 Pedestrian gathering detected by CAV}

\begin{figure}[t]
\centering
\subfigure[]{
\begin{minipage}[b]{0.47\textwidth}
\includegraphics[width=1\textwidth]{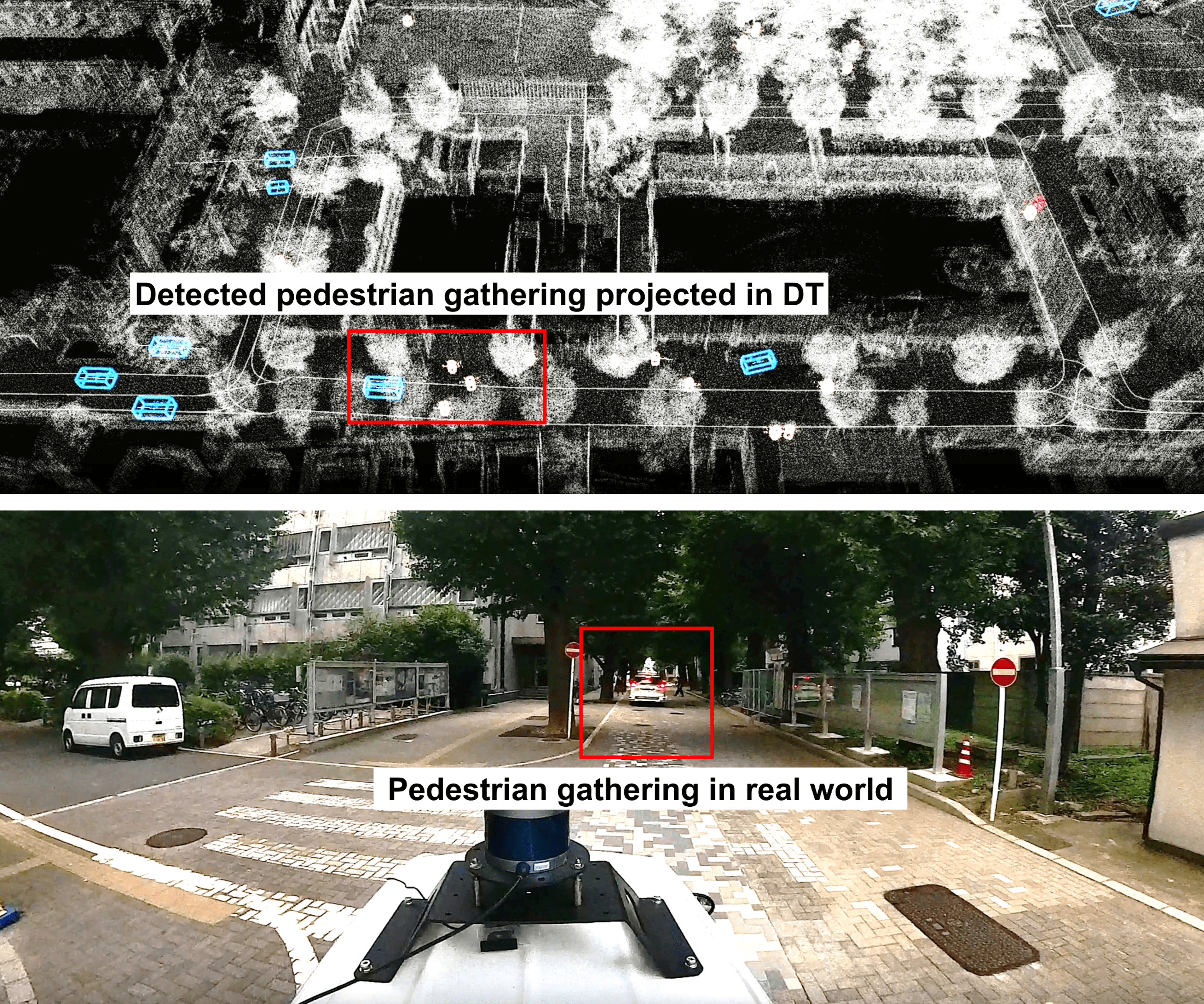} 
\end{minipage}
}
\subfigure[]{
\begin{minipage}[b]{0.47\textwidth}
\includegraphics[width=1\textwidth]{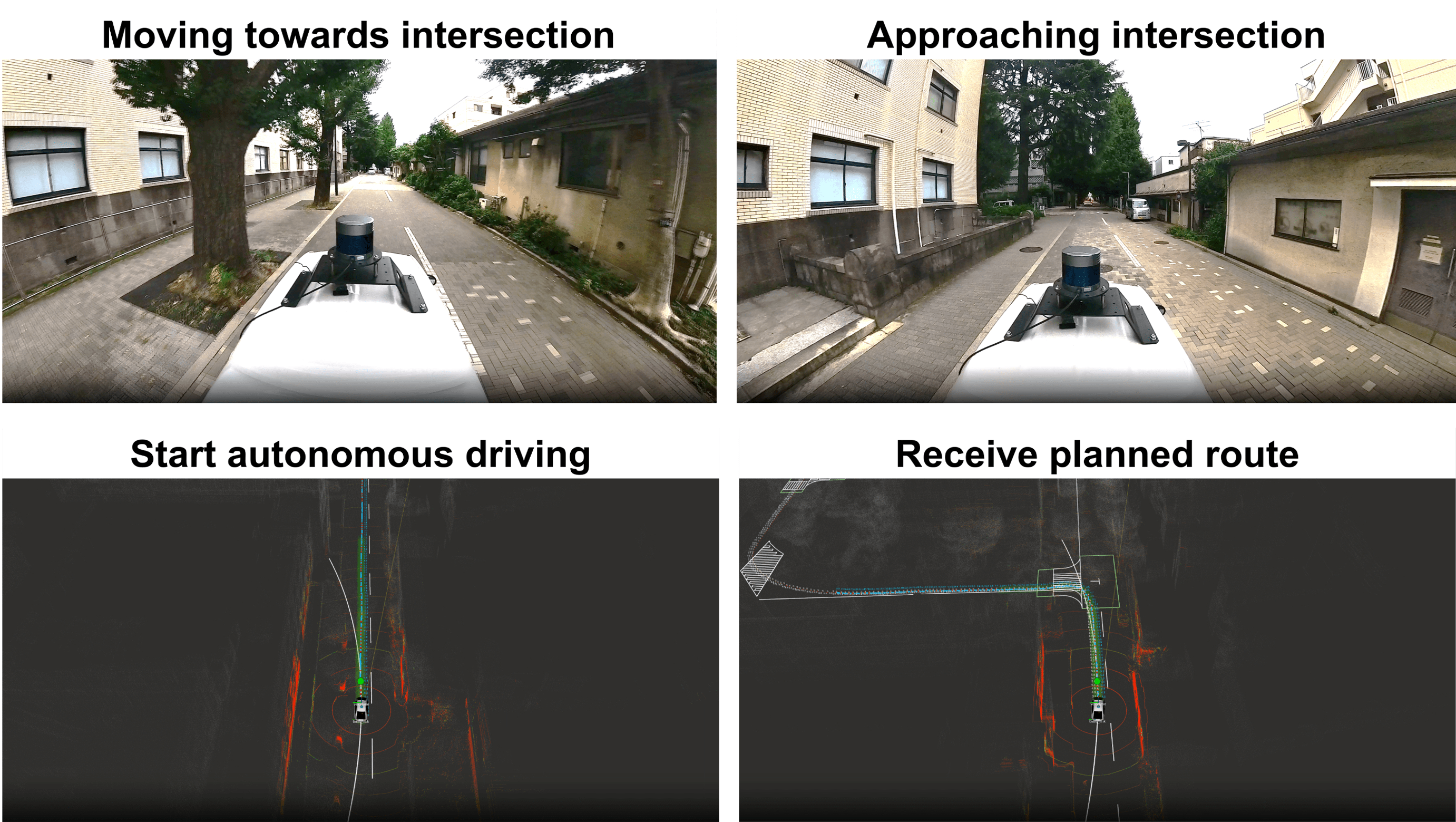} 
\end{minipage}
}
\caption{Illustration of traffic DT and autonomous driving operation results in Case \#1. (a): visualization of traffic DT and real-world image showing the congestion caused by pedestrian gathering, (b) car-following camera view and ego vehicle Rviz view showing the executed route.}
\label{fig: result1}
\end{figure}

\begin{figure}[t]
\centering
\subfigure[]{
\begin{minipage}[b]{0.47\textwidth}
\includegraphics[width=1\textwidth]{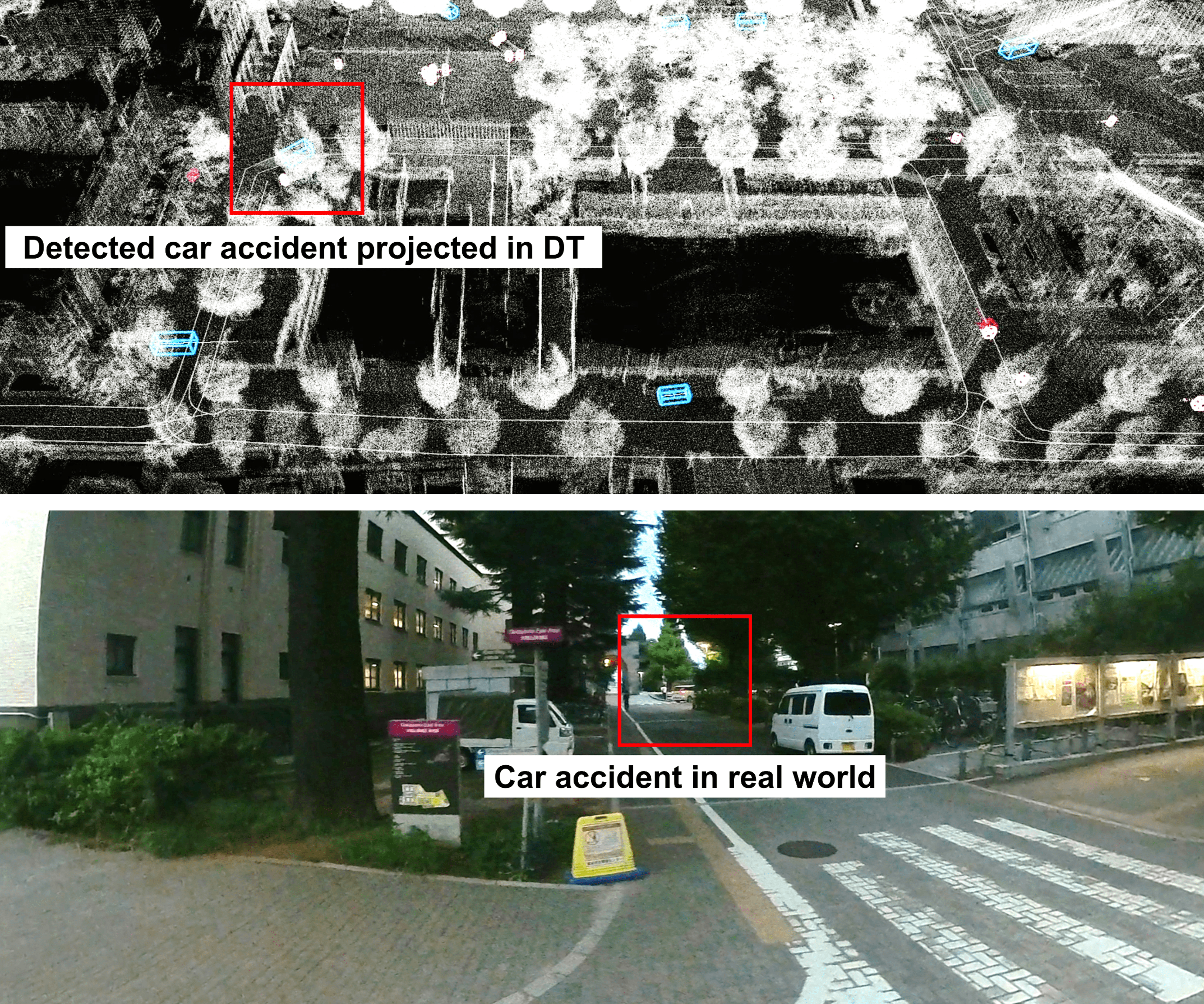} 
\end{minipage}
}
\subfigure[]{
\begin{minipage}[b]{0.47\textwidth}
\includegraphics[width=1\textwidth]{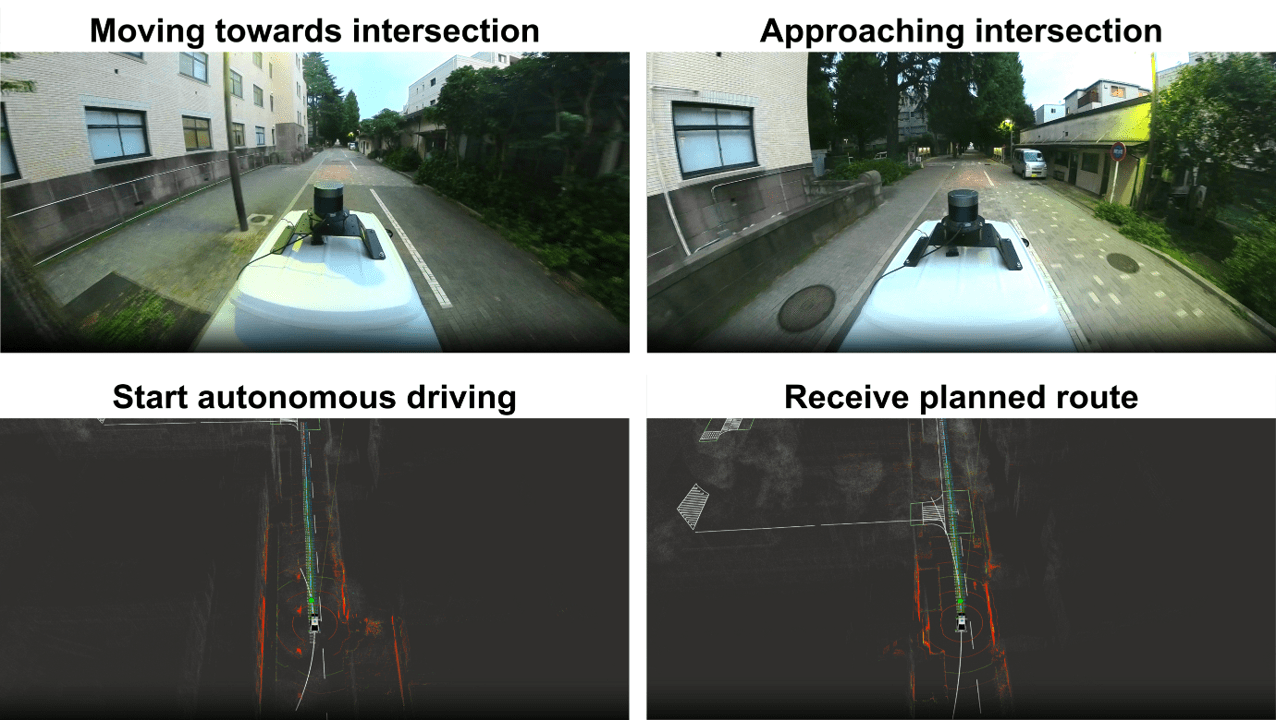} 
\end{minipage}
}
\caption{Illustration of traffic DT and autonomous driving operation results in Case \#2. (a): visualization of traffic DT and real-world image showing the congestion caused by pedestrian gathering, (b) car-following camera view and ego vehicle Rviz view showing the executed route.}
\label{fig: result2}
\end{figure}

Fig.~\ref{fig: result1} shows the results of SMDT-based route planning process in Case \#1, Fig.~\ref{fig: result1}(a) gives a view of traffic DT on cloud server, where the pink bounding boxes represent detected pedestrians. It can be seen that there are three pedestrians walking on the roadway of Route \#1. This information is then captured by CAV \#2, which is the first CAV entering the road network, and subsequently uploaded to the cloud. A car-following camera mounted on CAV \#1 to record its driving process, also captures this scene, where CAV \#2 halts in front of the gathering crowd. Fig.~\ref{fig: result1}(b) shows the autonomous driving process of CAV \#1 from the perspectives of the car-following camera and the Autoware visualization tool Rviz, respectively. During the initialization phase, CAV \#1 is proceeding straight, poised to enter the road network. When CAV \#1 arrives at the predetermined threshold distance $S_{\textrm{thre}}$ from the intersection area, it will start sending its position to the cloud through V2C communication to notify its impending entry into the network and requests route planning service. Then the cloud sends the Route \#2 's Waypoints file to CAV \#1 to help it avoid pedestrian gathering area and navigate it to destination. Consequently, CAV \#1 executes the planned route to turn left at the intersection and finally arrives at the destination. In this experiment, the threshold distance $S_{\textrm{thre}}$ is determined according to equation (5) with maximum vehicle speed:

\begin{equation}
    S_{\textrm{thre}} = 0.164~v_{\textrm{max}}^2 = 5.062~m
\end{equation}

\subsubsection{Case \#2 Traffic accident detected by RSU}

Similar to Case \#1, Fig.~\ref{fig: result2}(a) shows the traffic DT, where a parked car can be found on Route \#2, represented by a blue bounding box. Under such circumstances, the cloud system would deduce that a car accident has occurred on Route \#2. In Fig.~\ref{fig: result1}(b), we also give a real-world image from car-following camera on CAV \#1 that records the parking car and executed route in Rviz. In this case, the cloud sends Route \#1 's Waypoints to CAV \#1 to help it avoid the car accident on Route \#2.

The case study demonstrates the effectiveness of SMDT-based route planning for autonomous vehicles: \textit{(i)} the system's ability to detect and respond to real-time events, such as pedestrian gatherings and traffic accidents, and \textit{(ii)} reliable route planning service to achieve the control on vehicle level. This underscores the system's potential for real-world application in dynamic traffic environments. However, challenges for future development mainly include safety and robustness issues. Safety concerns primarily arise from the reliance on the V2X communication network and the accuracy of traffic monitoring, especially in unpredictable urban environments with high CAV density. Robustness is challenged by varying environmental conditions and sensor limitations. To address these, future work could: \textit{(i)} enhance V2X network resilience by implementing advanced encryption and cybersecurity protocols and deploying decentralized communication networks, such as blockchain technology, \textit{(ii)} expand the variety of sensors and integrate sensor fusion techniques to provide a more comprehensive and fail-safe approach to environmental perception, \textit{(iii)} explore the use of big traffic data analytics and deploy advanced machine learning algorithms to analyze vast amounts of traffic data more efficiently and accurately, and \textit{(iv)} develop more sophisticated decision-making models that consider a wider range of variables, such as road surface and weather conditions.

\section{Large-Scale Traffic Simulation and System Evaluation}
To evaluate the effectiveness of proposed CAV navigation system and crucial KPIs of implemented SMDT platform, a set of simulations and experiments are conducted. SUMO is utilized as a large-scale traffic simulator, where we can import test areas from the open street map (OSM), then generate random vehicles and control them with TraCI4Matlab \cite{acosta2015traci4matlab}. Subsequently, we can deploy the cooperative event-triggered route planning algorithm in MATLAB to realize real-time path planning for CAV users. Finally, we empirically measured several metrics to validate whether our implemented platform meets the requirements posed by the CAV navigation system.

\subsection{Efficiency and Safety Improved by CAV Navigation System}

\begin{figure}[t]
    \centerline{\includegraphics[width=0.43\textwidth]{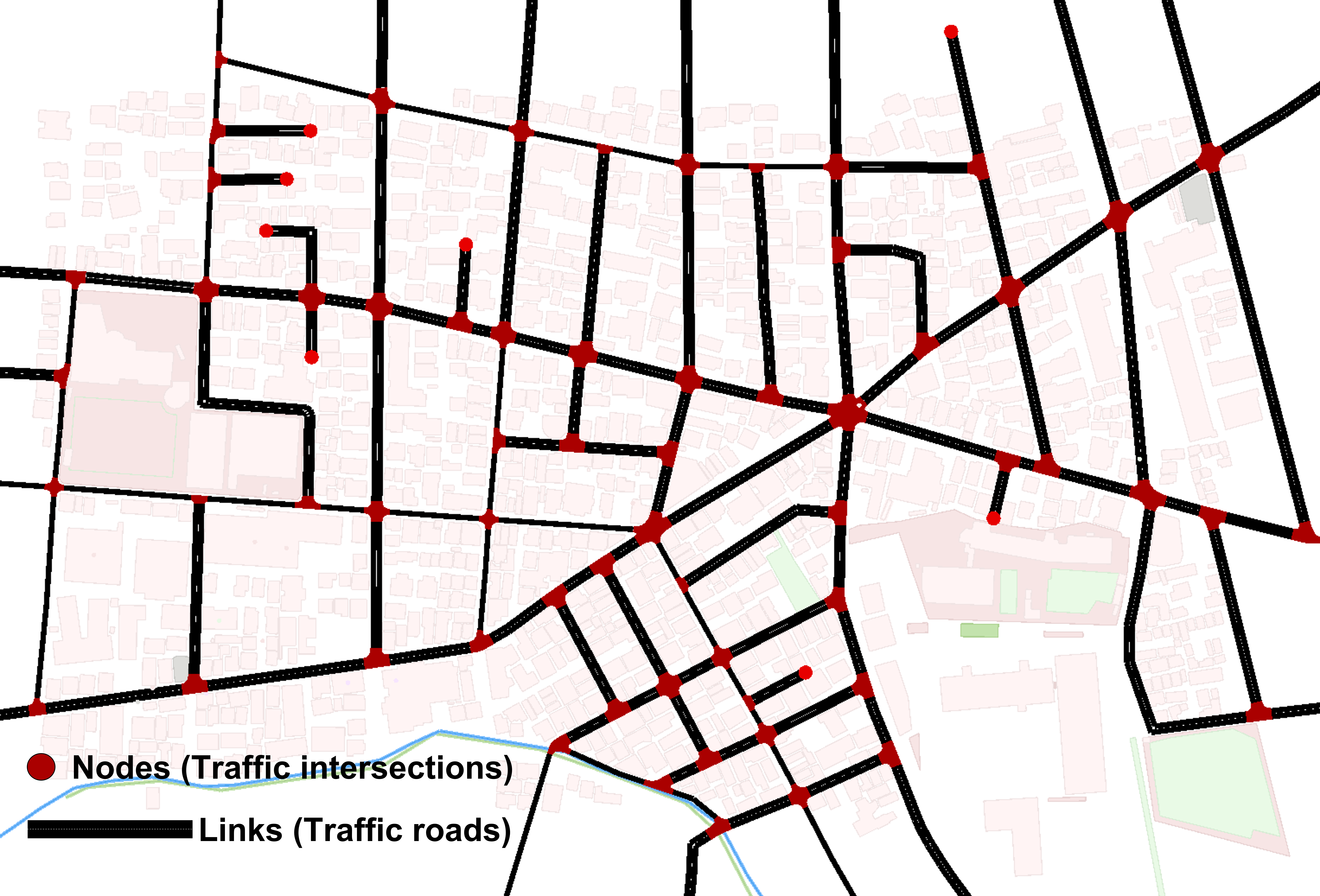}}
    \caption{Traffic network of SUMO simulation in Tokyo, Japan.}
    \label{fig: sumonet}
\end{figure}

\begin{table}[t]
\centering
\label{tab: para}
\caption{Simulation Parameters}
\begin{tabular}{c|c|c}
\hline
Parameters                          & Description           & Value    \\ \hline
$\Delta T$                          & Sampling time         & 1 s       \\
$T_{\textrm{sim}}$                  & Total simulation time & 600 s     \\
$M$                                 & Amount of nodes       & 90       \\
$M_{\textrm{link}}$                    & Amount of links       & 504      \\
$N_{\textrm{vel}}$                  & Amount of vehicles    & 300      \\
$N_{{\textrm{user}}}$               & Amount of CAV users   & 10 - 100 \\
$P_{{\textrm{user}}}$               & Proportion of CAV users   & 3.3 - 33.3\% \\
$E$                                 & Amount of traffic events    & 0 - 10   \\ \hline
\end{tabular}
\end{table}

In SUMO-based traffic simulation, we import an urban traffic area located in Tokyo as illustrated in Fig.~\ref{fig: sumonet}. The speed limit for each road remains consistent with the default values in OSM, and every intersection is designed to be non-signalized. During the real-time simulation process, at each sampling time, a random number of vehicles enter the traffic network with both their origins and destinations being randomized. A certain proportion of these vehicles are designated as CAV users, for whom we provide real-time route planning services. In contrast, the remaining vehicles are classified as unconnected vehicles. Lacking dynamic traffic information, they opt for the shortest path between their origin-destination (OD) pairs. Throughout the simulation, a certain number of traffic events, i.e., \emph{pedestrian gathering} and \emph{traffic accidents}, occur at random places in the traffic network. Table VI summarizes the main parameters used in this simulation.

\begin{figure}[t]
\centering
\subfigure[]{
\begin{minipage}[b]{0.215\textwidth}
\includegraphics[width=1\textwidth]{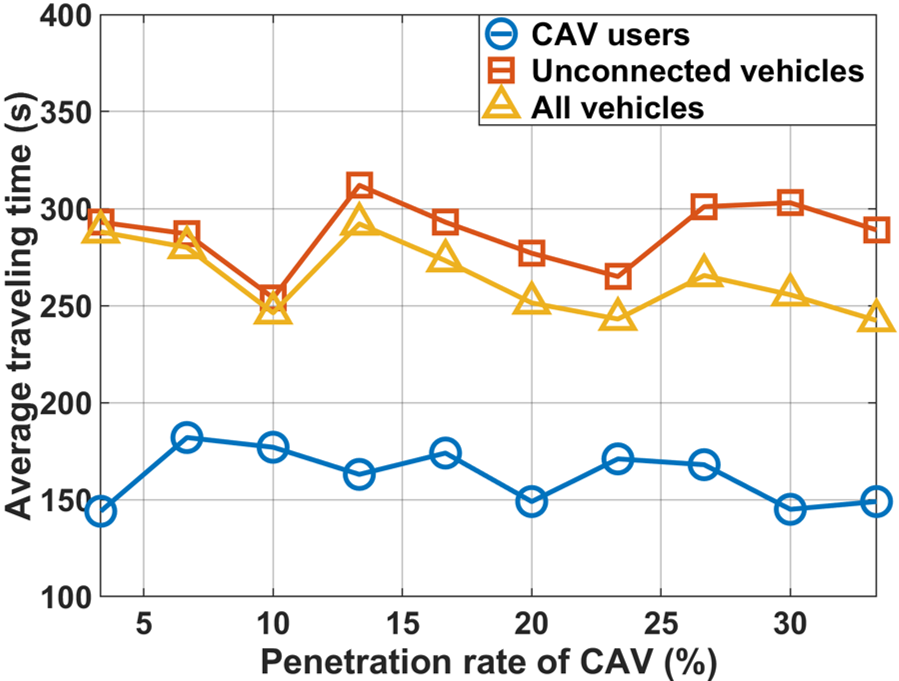} 
\end{minipage}
}
\subfigure[]{
\begin{minipage}[b]{0.215\textwidth}
\includegraphics[width=1\textwidth]{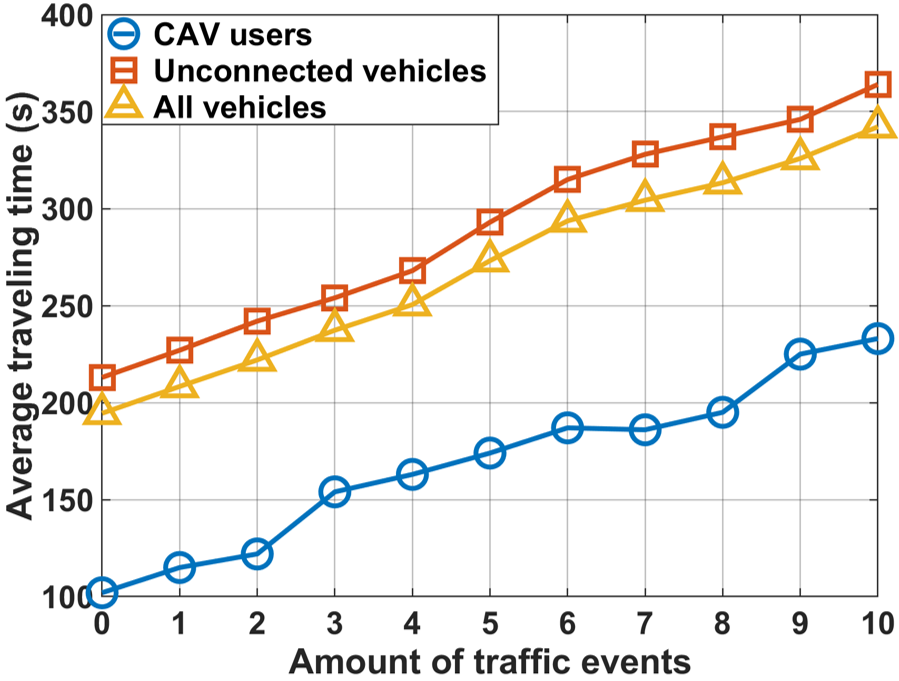} 
\end{minipage}
}
\subfigure[]{
\begin{minipage}[b]{0.215\textwidth}
\includegraphics[width=1\textwidth]{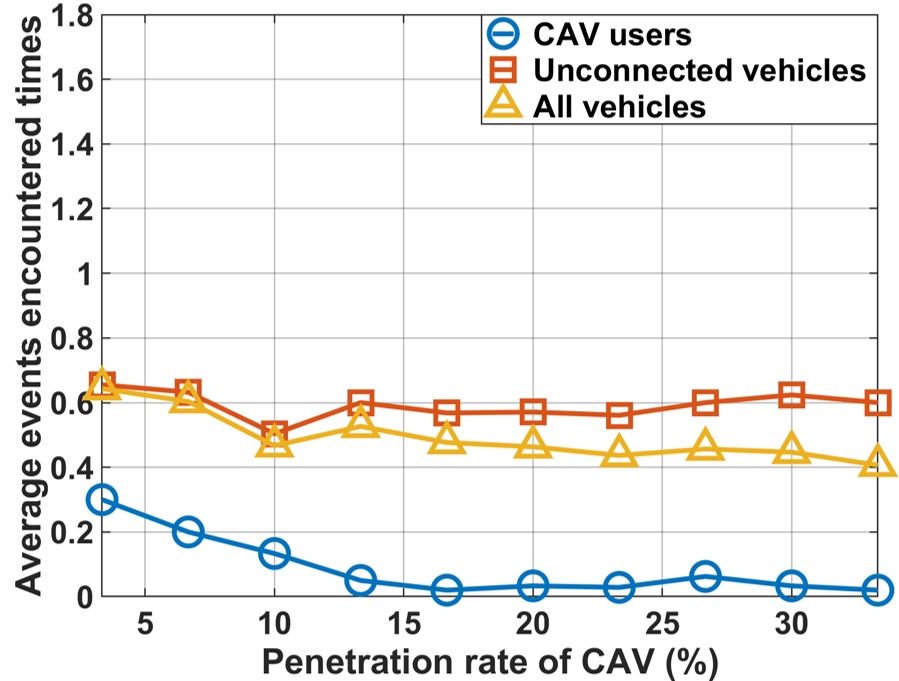} 
\end{minipage}
}
\subfigure[]{
\begin{minipage}[b]{0.215\textwidth}
\includegraphics[width=1\textwidth]{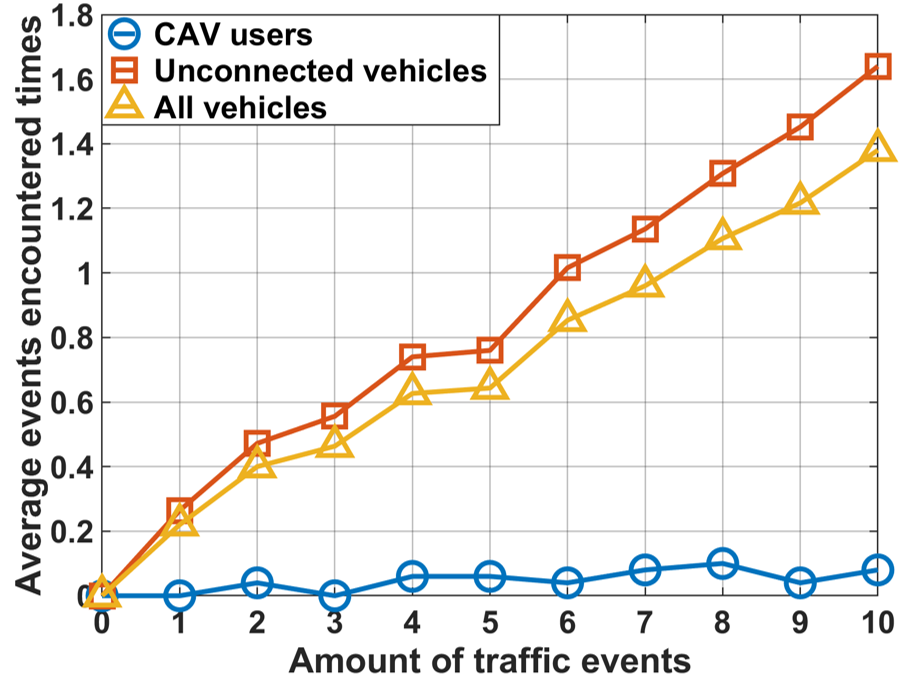} 
\end{minipage}
}
\caption{Simulation results of traffic efficiency and road safety: (a) average travel time with increasing penetration rate of CAV $P_{\textrm{user}}$ ($E=5$), (b) average travel time with increasing number of traffic events $E$ ($P{\textrm{user}} =16.7$\%), (c) average event encountered times with increasing penetration rate of CAV $P_{\textrm{user}}$ ($E=5$), (d) average event encountered times with increasing number of traffic events $E$ ($P_{\textrm{user}} =16.7$\%).}
\label{fig: simures1}
\end{figure}

Fig.~\ref{fig: simures1} shows the improved traffic efficiency and road safety with our CAV navigation system. We employ the frequency of encounters with traffic events as an evaluation metric for road safety since such extreme events could pose latent threats and safety concerns for vehicles passing by. In Fig.~\ref{fig: simures1}(a), As the number of events increases, both CAV users and unconnected vehicles experience longer driving time to traverse the network. This is predominantly due to that congestion caused by traffic events might propagate to other road segments. Nevertheless, the travel time for CAV users is significantly smaller than that of unconnected vehicles. In Fig.~\ref{fig: simures1}(b), it can be observed that, given a constant number of events, with the number of CAV users increasing, the efficiency of CAV users cannot be improved. However, the travel time for the overall traffic consistently shows a declining trend, where some fluctuations are attributed to the random selection of OD pairs and the random occurrence locations of traffic events. As can be discerned from Fig.~\ref{fig: simures1}(c), with the occurrence of an increasing number of events, it can be seen that the frequency with which CAV users encounter these events does not show an upward trend. This suggests that the \emph{event-triggered mechanism} in our route planning strategy is effective in assisting users to circumvent such events. As depicted in Fig.~\ref{fig: simures1}(d), with an increasing penetration rate of CAV users, the safety of both CAV users and the overall traffic can be improved, manifesting as a reduction in the frequency of encounters with traffic events.

\subsection{System Evaluation Considering Communication Issues}

In DT modeling, the packet delivery rate (PDR) for communication from RSUs to the cloud server is very critical and important, since significant packet loss may lead to incorrect planning decisions. When only processed sensing data (i.e., detection and tracking results) are transmitted, the PDR can approach $100\%$. However, when simultaneously uploading other levels of data, such as raw LiDAR point clouds, the PDR slightly decreases to approximately $99.53\%$, which still satisfies the reliability criterion for SSMS, which mandates a PDR greater than $95\%$. As for the process of providing route planning service, V2C communication is built upon the HTTP protocols. Since HTTP operates atop the Transmission Control Protocol (TCP) at the application layer and TCP ensures data delivery through error detection and retransmissions, the reliability of data transfer under the HTTP protocol can be ensured in stable network conditions.

\begin{table}[t]
\centering
\label{tab:latency}
\caption{Communication Requirements and Performances}
\begin{tabular}{cccc}
\hline
                                                                           & Max. (ms) & Min. (ms) & Mean (ms)   \\ \hline
\multicolumn{4}{l}{\textbf{DT modeling process}}                                                                                  \\ \hline
I2C comm.                                                                  & 1.74      & 1.10      & 1.37              \\
\begin{tabular}[c]{@{}c@{}}Edge comp.\\ (detection)\end{tabular}           & 153.41    & 70.01    & 106.23                       \\ \hline
\multicolumn{4}{c}{$T_{\textrm{dt}}$ Total Max.: 155.15~ms}                                                                     \\ \hline
\multicolumn{4}{l}{\textbf{Route planning service}}                                                                               \\ \hline
V2C comm.                                                                  & 42.13     & 20.16     & 32.30            \\
\begin{tabular}[c]{@{}c@{}}Cloud comp.\\ (traffic monitoring)\end{tabular} & 56.29     & 42.72     & 45.07                       \\
\begin{tabular}[c]{@{}c@{}}Cloud comp.\\ (route planning)\end{tabular}     & 201.07    & 173.27    & 183.68                      \\
\begin{tabular}[c]{@{}c@{}}Edge comp.\\ (NDT localization)\end{tabular}    & 10.13     & 2.56      & 6.14                        \\
\begin{tabular}[c]{@{}c@{}}Edge comp.\\ (route loading)\end{tabular}       & 500.97    & 501.35    & 501.18                      \\ \hline
\multicolumn{4}{c}{$T_{\textrm{svc}}$ Total Max.: 810.59$< 0.164~v_{\textrm{max}} = 911.11$~ms}                                                                   \\ \hline
\end{tabular}
\end{table}

Some crucial communication and computational latency existing in our system are also measured, as summarized in Table VII. It can be found that both the maximum One-way I2C ($1.74$~ms) and V2C communication latency ($42.1$~ms) can meet the requirements proposed by 3GPP, which are $82.6\%$ below the threshold of the max E2E latency for SSMS (less than $10$~ms), and $57.9\%$ below the threshold for information sharing (less than $100$~ms). 

The total latency for \emph{traffic DT modeling} $T_{\textrm{dt}}$ is primarily constituted by the latency derived from edge computing $T_{\textrm{dec}}$ and the one-way I2C communication $T_{\textrm{I2C}}$, both of which are measured through experiments. Given that the quantity of objects within the LiDAR's sensing range has a large impact on the detection duration, we have conducted measurements of the latency spanning the initiation of the LiDAR to the computation of the detection outcomes many times at various intervals throughout the day. The edge computing exhibited a maximal latency of $153.41$~ms, whereas the overall latency for the \emph{traffic DT modeling} reached a peak of $155.15$~ms, according to equation (6). 

As for the \emph{route planning service} latency, we separately measure V2C communication delay $T_{\textrm{V2C}}$, cloud computing time $T_{\textrm{cloud}}$ for both traffic monitoring and route planning, and edge computing time for NDT-based localization and route loading/execution. Only the cloud computational time is measured with SUMO-based large-scale simulation, instead of with PoC experiment, because cloud computing in PoC just selects a route from two options, without searching for the fastest route. Based on equation (8), the maximal total latency of the \emph{route planning service} reaches $810.59$~ms, which complies with the requirement of CAV user receiving and executing the routing command prior to entering the intersection area. Additionally, the latency of \emph{traffic DT modeling} is also significantly lower than that of \emph{route planning service}. 

In the real-world experiment, the frequency for LiDAR-based object detection is set at 30~Hz, which is regarded as the frequency of \emph{traffic DT modeling}. Route planning, on the other hand, is executed on-demand, triggered when the CAV approaches an intersection within the threshold distance $S_{\textrm{thre}}$. Considering the update frequency and its associated delays, in extreme cases, the latency for \emph{traffic DT modeling} will reach up to 188.48~ms, which is also much lower than that of \emph{route planning service}. For simulations, the cloud computing capabilities operate with a sampling time of 1~s, which is larger than the measured maximum latency of \emph{route planning service}. This setup ensures that cloud services can be completed in real time within each sampling period, demonstrating the cloud's ability to process critical information promptly and efficiently. Taking into account the three crucial aspects of reliability, latency, and frequency, the traffic DT can be regarded as real-time and effective in the context of CAV navigation system.

\section{Conclusion}
In this study, we have proposed an SMDT platform to offer cloud services for CAV users. By design, this system integrates cloud and edge computing for the establishment of traffic DT. Based on the SMDT platform, we have designed a CAV navigation system that can utilize real-time traffic data and plan routes for CAVs, helping them circumvent dynamic traffic events and improving their efficiency and road safety. In our Smart Mobility Field, we have demonstrated that the designed SMDT-based CAV navigation system can realize effective and real-time navigation in response to traffic events. Future work will aim to expand this platform city-wide, incorporating advanced predictive algorithms for enhanced traffic modeling and exploring its integration within smart city frameworks.

\bibliographystyle{IEEEtran.bst}
\bibliography{bibliography}

\begin{IEEEbiography}
[{\includegraphics[width=1in,height=1.25in,clip,keepaspectratio]{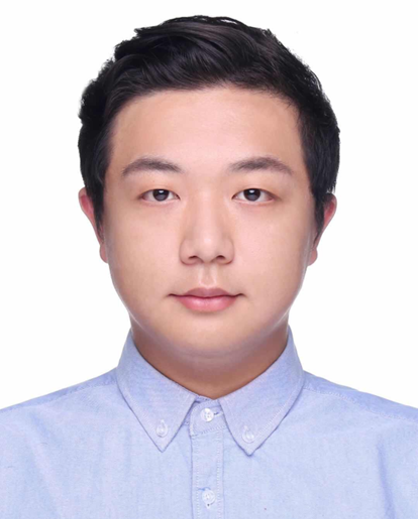}}]{Kui Wang} (Student Member, IEEE) received the B.E. degree in mechanical engineering from the Beijing University of Posts and Telecommunications, Beijing, China, in 2018, and the M.E. degree in vehicle engineering from the KTH Royal Institute of Technology, Sweden, in 2021. He is currently working toward the Ph.D. degree in the Department of Electrical and Electronic Engineering, Tokyo Institute of Technology, Tokyo, Japan. His current research interests include smart mobility, autonomous driving, digital twin, and federated learning.
\end{IEEEbiography}

\begin{IEEEbiography}[{\includegraphics[width=1in,height=1.25in,clip,keepaspectratio]{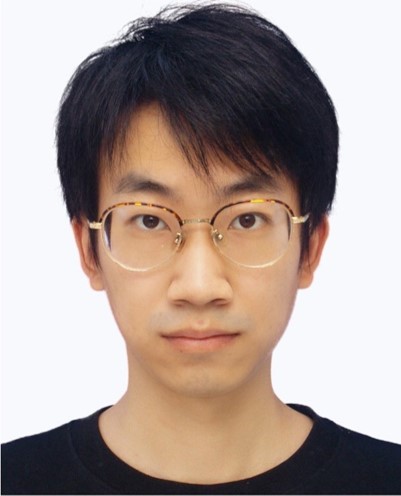}}]{Zongdian Li} (Member, IEEE) received the B.E. degree in communication engineering from Beijing University of Posts and Telecommunications (BUPT), Beijing, China, in 2018, and the M.E. and Ph.D. degrees in electrical and electronic engineering from Tokyo Institute of Technology (Tokyo Tech), Tokyo, Japan, in 2020 and 2023, respectively. Since October 2023, he has been an Assistant Professor with the Department of Electrical and Electronic Engineering, School of Engineering, Tokyo Tech. His research interests include software-defined networking (SDN), vehicle-to-everything (V2X) communications, digital twins, and autonomous driving. He is a member of IEICE.
\end{IEEEbiography}

\begin{IEEEbiography}[{\includegraphics[width=1in,height=1.25in,clip,keepaspectratio]{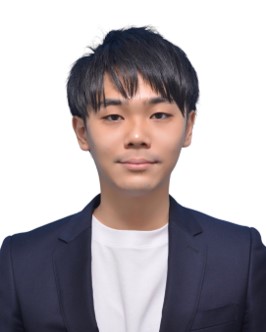}}]{Kazuma Nonomura} (Student Member, IEEE) received the B.E. degree in electrical and electronic engineering from the Tokyo Institute of Technology, Tokyo, Japan, in 2022. He is currently working toward the M.E. degree in the Department of Electrical and Electronic Engineering, Tokyo Institute of Technology, Tokyo, Japan. His current research interests include smart mobility, autonomous driving, digital twin, and reinforcement learning.
\end{IEEEbiography}

\begin{IEEEbiography}[{\includegraphics[width=1in,height=1.25in,clip,keepaspectratio]{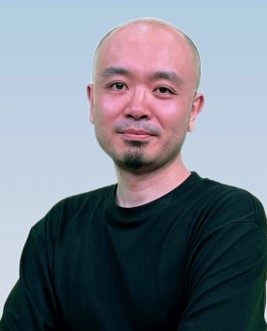}}]{Tao Yu} (Member, IEEE) received the M.E. degree from the Communication University of China (CUC), Beijing, China in 2010, and the Dr.Eng. degree from the Tokyo Institute of Technology (Tokyo Tech), Tokyo, Japan in 2017. From 2017 to 2022, he worked as a postdoctoral researcher in Sakaguchi Lab, Dept. Electr. Electron. Eng., Tokyo Tech. From 2022, he joined the Tokyo Tech Academy for Super Smart Society where he is currently a specially appointed associate professor. His research interests include UAV communication, V2X, sensor networks, localization, antenna, smart mobility, and building energy management. He is member of IEEE and IEICE.
\end{IEEEbiography}

\begin{IEEEbiography}[{\includegraphics[width=1in,height=1.25in,clip,keepaspectratio]{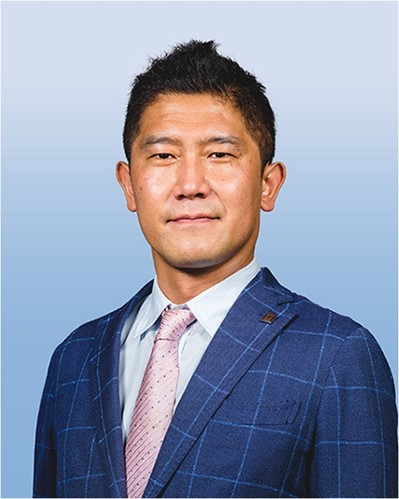}}]{Kei Sakaguchi} (Senior Member, IEEE) received the M.E. degree in information processing from Tokyo Institute Technology, Tokyo, Japan, in 1998, and the Ph.D. degree in electrical and electronics engineering from Tokyo Institute Technology, Tokyo, Japan, in 2006. He is currently working with the Tokyo Institute of Technology in Japan, as the Dean with Tokyo Tech Academy for Super Smart Society and as a Professor with the School of Engineering. At the same time, he is working for oRo Co., Ltd. in Japan as the outside Director. His current research interests include 5G cellular networks, millimeter-wave communications, wireless energy transmission, V2X for automated driving, and super smart society. He was the recipient of the Outstanding Paper Awards from SDR Forum and IEICE, in 2004 and 2005, respectively, and three Best Paper Awards from IEICE communication society in 2012, 2013, and 2015. He was also the recipient of the Tutorial Paper Award from IEICE communication society in 2006. He is a fellow of IEICE.
\end{IEEEbiography}

\begin{IEEEbiography}[{\includegraphics[width=1in,height=1.25in,clip,keepaspectratio]{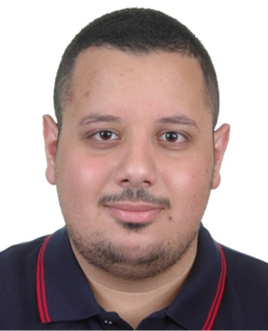}}]{Omar Hashash} (Student Member, IEEE) received his B.E. in Communications and Electronics Engineering from Beirut Arab University, Lebanon in 2019 and his M.E. in Electrical and Computer Engineering from the American University of Beirut, Lebanon in 2021. He is currently a Ph.D. student at the Electrical and Computer Engineering Department at Virginia Tech. His research interests include 6G wireless networks, digital twins, metaverse, edge intelligence, and generalizable AI. He has recently published one of the first works that explore the synergy between wireless, computing, and AI techniques that can come together to support massive digital twinning of physical systems in the metaverse.
\end{IEEEbiography}

\begin{IEEEbiography}[{\includegraphics[width=1in,height=1.25in,clip,keepaspectratio]{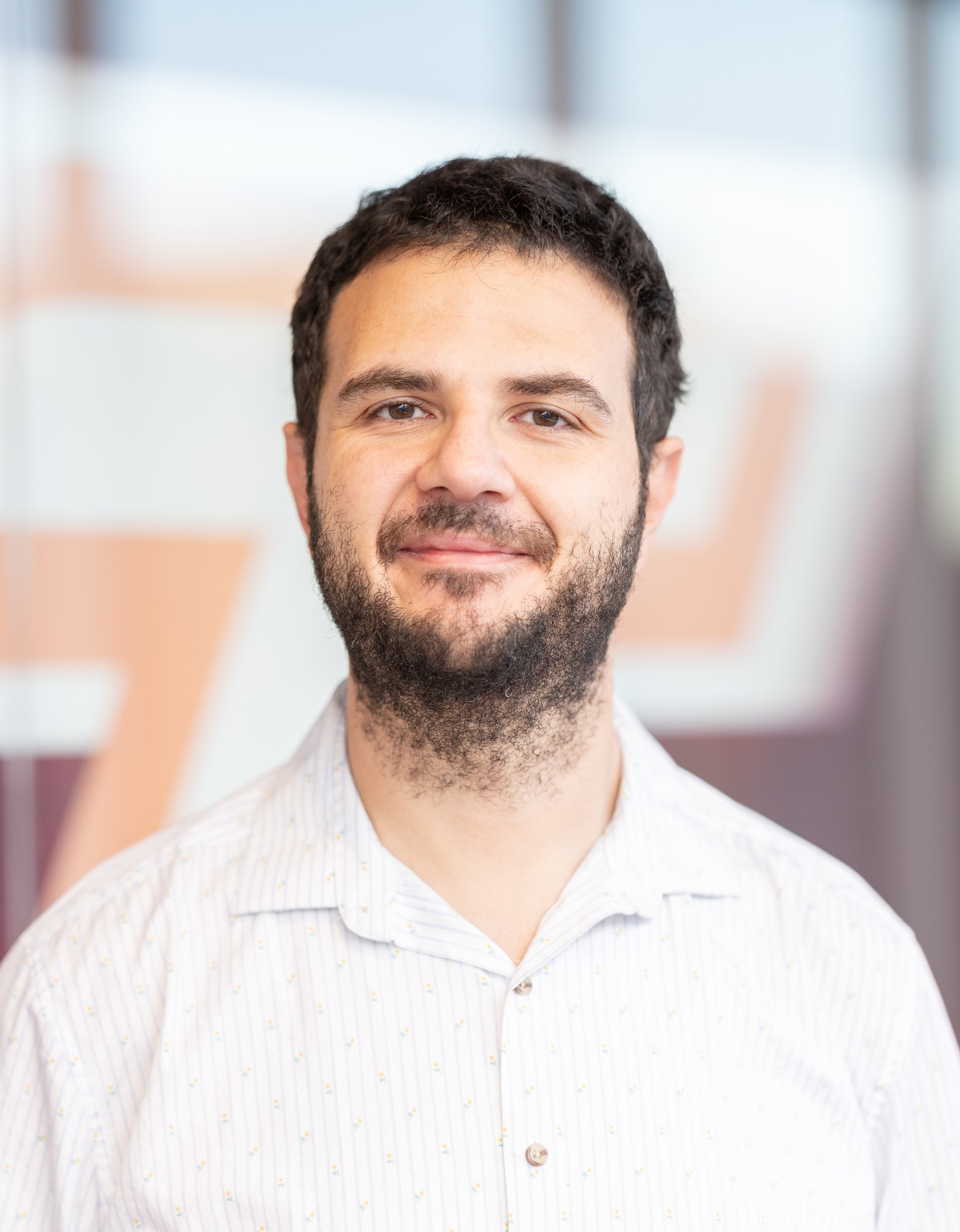}}]{Walid Saad} (Fellow, IEEE) received his Ph.D degree from the University of Oslo, Norway in 2010. He is currently a Professor at the Department of Electrical and Computer Engineering at Virginia Tech, where he leads the Network sciEnce, Wireless, and Security (NEWS) laboratory. His research interests include wireless networks (5G/6G/beyond), machine learning, game theory, security, UAVs, semantic communications, cyber-physical systems, and network science. Dr. Saad is a Fellow of the IEEE. He is also the recipient of the NSF CAREER award in 2013, the AFOSR summer faculty fellowship in 2014, and the Young Investigator Award from the Office of Naval Research (ONR) in 2015. He was the (co-)author of twelve conference best paper awards at IEEE WiOpt in 2009, ICIMP in 2010, IEEE WCNC in 2012, IEEE PIMRC in 2015, IEEE SmartGridComm in 2015, EuCNC in 2017, IEEE GLOBECOM (2018 and 2020), IFIP NTMS in 2019, IEEE ICC (2020 and 2022), and IEEE QCE in 2023. He is the recipient of the 2015 and 2022 Fred W. Ellersick Prize from the IEEE Communications Society, of the IEEE Communications Society Marconi Prize Award in 2023, and of the IEEE Communications Society Award for Advances in Communication in 2023. He was also a co-author of the papers that received the IEEE Communications Society Young Author Best Paper award in 2019, 2021, and 2023. Other recognitions include the 2017 IEEE ComSoc Best Young Professional in Academia award, the 2018 IEEE ComSoc Radio Communications Committee Early Achievement Award, and the 2019 IEEE ComSoc Communication Theory Technical Committee Early Achievement Award. From 2015-2017, Dr. Saad was named the Stephen O. Lane Junior Faculty Fellow at Virginia Tech and, in 2017, he was named College of Engineering Faculty Fellow. He received the Dean's award for Research Excellence from Virginia Tech in 2019. He was also an IEEE Distinguished Lecturer in 2019-2020.  He has been annually listed in the Clarivate Web of Science Highly Cited Researcher List since 2019. He currently serves as an Area Editor for the IEEE Transactions on Network Science and Engineering and the IEEE Transactions on Communications. He is the Editor-in-Chief for the IEEE Transactions on Machine Learning in Communications and Networking.
\end{IEEEbiography}

\vfill

\end{document}